\documentclass{aa}
\usepackage{graphicx}

\begin{document}

        \title{Models for the soft X-ray emission of post-outburst classical novae}

        \author{G. Sala \and M. Hernanz}

        \offprints{M. Hernanz}

        \institute{Institut d'Estudis Espacials de Catalunya and 
Institut de Ci\`encies de l'Espai (CSIC).\\Campus UAB, Facultat de Ci\`encies, 
Torre C5-parell, 2a planta. E-08193 Bellaterra (Barcelona), 
Spain.\\ \email{sala@ieec.uab.es, hernanz@ieec.uab.es}}

        \date{Received ... /accepted ...}

        \abstract{A hydrostatic and stationary white dwarf 
envelope model has been developed for the study of the post-outburst phases 
of classical novae and their soft X-ray emission. Several white dwarf 
masses and chemical compositions typical for classical novae have been considered. 
The results show that the luminosity, maximum effective temperature and envelope masses 
depend on the white dwarf mass and on the chemical composition. Envelope masses for 
which equilibrium solutions exist are pretty small ($\sim 10^{-7}-10^{-6}\rm M_{\odot}$), 
thus leading to a short duration of the soft X-ray emitting phase 
of classical novae, in agreement with most of the observations.
The models presented provide a useful tool for the determination of the 
white dwarf properties from observable parameters in the X-ray range.
}

        \titlerunning{Envelope models for post-outburst novae}
        \maketitle
%

\section{Introduction}

Classical nova outbursts are caused by the explosive burning of hydrogen 
on the surface of the white dwarf of a cataclysmic variable, 
a close binary system with mass transfer from a main sequence star (secondary) 
to a white dwarf (primary).
Hydrodynamic models show that, after $10^{-5}-10^{-4}\rm M_{\odot}$
of H-rich material are transferred from the secondary star to the white
dwarf, ignition in degenerate conditions takes place in the accreted
envelope and a thermonuclear runaway is initiated (Starrfield \cite{sta89}). 
As a consequence of the thermonuclear runaway, the envelope expands and causes the
brightness of the star to increase by $\sim$10 magnitudes, reaching
maximum luminosities up to $\sim10^{5}\rm L_{\odot}$. A fraction
of the envelope is ejected at large velocities (hundreds or thousands
of  km/s), while a part of it returns to hydrostatic
equilibrium with steady nuclear burning. 
Hydrodynamic models also show that a powerful outburst is only possible if
the accreted matter (with solar abundances) is enhanced by some mixing
mechanism with CNO nuclei from the degenerate core. Analyses of nebular
emission lines from classical novae show indeed ejecta enhanced in
CNO nuclei compared to cosmic abundances and, in some cases, enhancement
in Ne, Mg and other heavier elements.

The presence of a remaining hydrogen burning envelope on the post-outburst 
white dwarf was already pointed out after the first
hydrodynamic simulations indicated that
only a fraction of the hydrogen rich accreted material is ejected  
(Starrfield et al. \cite{sta74}).
The rest of the envelope is expected to return to hydrostatic equilibrium
with steady hydrogen burning, powering a source of constant bolometric luminosity.
As the ejected envelope expands, the photosphere receedes, and inner and hotter 
layers become visible, shifting the spectrum from optical to ultraviolet, and finally 
to soft X-rays. This picture could explain the persistence of the
high temperature ultraviolet emission after the optical decline, 
detected for the first time with OAO-2 
in Nova Serpentis 1970 (Gallagher \& Code \cite{gal73})
and in other novae with IUE (1978-1996).

Without a detailed model for the post-outburst envelope, the duration of the constant
bolometric luminosity phase was estimated as the nuclear
timescale of the mass left on the white dwarf surface (see for instance 
table 1 in Gehrz et al. \cite{geh98}). Since the outburst
ejects a fraction of the accreted envelope, the remaining envelope
mass was assumed to be of the same order of magnitude, 
which was estimated as the envelope mass needed to achieve 
the critical pressure at the base of the envelope to trigger 
the outburst (MacDonald \cite{mac83}), 
$\sim 10^{-4}-10^{-5}\rm M_{\odot}$ for white dwarf masses 
from 0.9 to 1.3 M$_{\odot}$ (in agreement with accreted 
envelope masses obtained with hydrodynamic models).
With the expected constant bolometric luminosity, $10^{4}-10^{5}\rm L_{\odot}$,
the nuclear time-scale during which the nova was expected to be a
bright soft X-ray source was of tens or hundreds of years.
According to these estimations, classical novae would be expected 
to shine as soft X-ray sources for decades or centuries. 
The spectrum of this super-soft X-ray emission is expected to be that 
of a hot white dwarf atmosphere (MacDonald \& Vennes \cite{mac91}).

The first detection of soft X-rays from a classical nova 
in outburst came from GQ Mus 1983, when observed
with EXOSAT in 1984, 460 days after optical maximum (\"Ogelman et al. \cite{oge84}).
Other novae were also detected some hundred days after optical
maximum (PW Vul, QU Vul, \"Ogelman et al. \cite{oge84}), but a major break-through was 
provided with the launch of the R\"ontgensatellit (ROSAT), 
which observed a total of 39 recent classical novae, less than 10 
years after the outburst. A systematic search for the X-ray emission from those 
post-outburst classical novae in the ROSAT
archival data was performed by Orio et al. (\cite{ori01}). Contrary to what
was expected, very few novae were discovered as X-ray emitters shortly
after their outburst. In fact only 3 novae showed a soft X-ray spectrum:
GQ Mus, V1974 Cyg and Nova LMC 1995. GQ Mus showed the longest constant
bolometric luminosity phase so far observed: its soft X-ray emission
lasted for 9 years, still less than predicted. V1974
Cyg, the best monitored soft X-ray nova, turned off only 18 months after the outburst. 
In the case of Nova LMC 1995, it was detected by ROSAT as a soft X-ray source three years
after outburst, and in a second observation with XMM-Newton in the
year 2000, five years after outburst.

X-ray observations give direct insights into the hydrogen burning layer
on the white dwarf surface after the outburst, but the presence of
the soft X-ray source can also be determined from
ultraviolet observations (Shore \cite{sho02}, Cassatella et al. \cite{cas04}).
The ultraviolet is the key spectral regime for studying the dynamics 
and opacity of the expanding shell, ionized
by the central X-ray source, and determine abundances, velocities, spatial structure
and mass of the ejecta. The ultraviolet emission
lines are indicators of the ionization and density structure of the
ejecta, and their variations reflect the evolution of the expanding
shell, but also that of the luminosity and effective temperature of
the central ionizing X-ray source.
This indirect indicator has been used to determine the turn-off of classical novae from
IUE observations (Shore et al. \cite{sho96}, Gonz\'alez-Riestra et al. \cite{gon98}, 
Vanlandingham et al. \cite{van01}).
In all cases, turn-off times were smaller than expected, and also
smaller than the 9 years of GQ Mus. 

The discrepancy between theory and observations pointed out that an
extra mass loss mechanism should be present shortly after the nova
outburst. Some mechanisms have been suggested, involving
the interaction of the ejecta with the envelope of the secondary and/or thick winds.
MacDonald et al. (\cite{mac85}) calculated that a fraction of the envelope could
be lost by dynamical friction of the secondary in a common envelope
phase following the nova outburst. Kato \& Hachisu (\cite{kat94})
calculated a sequence of optically thick wind solutions that simulated the decay
phase of the visual light curve of novae.

Nevertheless, the discrepancy between observations and theory 
arose from a comparison mainly based on 
the turn-off times. A model for post-outburst envelopes to
which X-ray observations could be compared was lacking. However, a
good number of works were devoted to the study of hot white dwarf
envelopes. Stellar evolution
models following the post AGB phase simulate a similar situation.
For population I stars (with X$_{\rm H}$=0.7), Paczy\'nski (\cite{pac70})
showed that after exhaustion of He in the core, stars of several initial
masses (between 3 and 7 M$_{\odot}$) follow a common track 
in the HR diagram with constant luminosity. Effective temperature increases along the track,
while hydrogen and helium shell burning reduce the envelope mass.
Paczy\'nski (\cite{pac70}) was also the first to find that luminosity is
directly proportional to the core mass during this phase, that lasts
as long as the envelope mass is larger than some critical value. When
this critical point is reached, the shell sources die out and the
degenerate core starts to cool down.

A number of studies have been devoted to accreting white dwarfs and
the development of thermal pulses or flashes on a hydrogen rich envelope.
Among them, Iben (\cite{ibe82}) and Iben \& Tutukov (\cite{ibe89}) found a core mass -
luminosity relation for accreting white dwarfs
with steady hydrogen burning in the accreted envelope 
(with X$_{\rm H}$=0.64) similar to Paczy\'nski's one. Iben (\cite{ibe82})
simulated the evolution of a white dwarf envelope without
accretion (or very small values of accretion rate) as a sequence of
steady-state models, and found that the constant luminosity phase
lasted for decades, in agreement with the nuclear timescales 
expected for the envelope masses of post-outburst classical novae
predicted by hydrodynamic models. 
The steady hydrogen burning solutions were
combined with an evolving quasi-static approximation to study the
thermal pulses occurring on accreting white dwarfs. Different points
of view of the problem can be found in Fujimoto (\cite{fuj82}) and Paczy\'nski (\cite{pac83}),
who developed simple approximations for the study of the hydrogen
burning envelopes stability, and the evolution of shell flashes for
different accretion rates. Taking into account the existing numerical
models for degenerate stars with hydrogen-burning shells, Iben \& Tutukov (\cite{ibe96})
pointed out that the core mass - luminosity relation was not unique,
and that it depended on the thermal history of the star. They compared
the models with data from symbiotic stars and cataclysmic variables,
exploring the effects of mechanisms not present in the models (such
as wind mass loss and mixing mechanisms) in the evolution of accreting
white dwarfs. Nevertheless, none of those works studied envelopes
with abundances typical for classical novae.
Tuchman \& Truran (\cite{tuc98}) studied the composition influence on the
core mass - luminosity relation and its implications for the turn-off
times of classical novae. They applied their relation to GQ Mus and
V1974 Cyg, and concluded that the envelope masses required to initiate
the thermonuclear runaway were larger than the envelope masses below
which stable models for steady hydrogen burning exist. This condition
led them to suggest that the remnant envelope masses in the post-outburst
stage could be significantly reduced by some kind of dynamical instability.

In this paper, a numerical model for white dwarf envelopes with
steady hydrogen burning is presented. The model provides
a composition dependent core mass - luminosity relation, and is
mainly aimed to simulate the conditions of the remaining hydrogen-rich
material left on the white dwarf after the nova outburst. Photospheric
properties are obtained for comparison with X-ray observations, and
their evolution is approximated as a sequence of steady-states. Section 2 describes
the envelope model, while the main results are presented in section 3. 
In section 4, analytical relations between the main global properties 
of the envelopes are given. The quasi-static evolution is 
presented in section 5, and a summary and discussion of the results
can be found in section 6.

\section{The white dwarf envelope model}

A grid of white dwarf envelope models with steady hydrogen burning
has been computed for five different core masses (from 0.9 to 1.3 M$_{\odot}$)
and four different chemical compositions, all of them corresponding to
realistic envelopes accreted onto white dwarfs in novae. Compositions are taken from  
Jos\' e  \& Hernanz (\cite{jh98}),
who developed hydrodynamic simulations of nova explosions with detailed
computation of the nucleosynthesis, and obtained results for the composition
of the ejecta compatible with observed abundances in many cases. As
already found from the first hydrodynamic models of nova outbursts,
their models showed that the material accreted from the secondary,
with solar abundances, needs to be enhanced with CNO nuclei by some
mixing mechanism between the envelope and the degenerate core.  
The compositions considered here correspond to CO white dwarf envelopes
with 50\% mixing of the solar accreted matter with the core material,
and to ONe white dwarf envelopes with 25\%, 50\% and 75\% mixing. 
For the calculation of the accreted envelope
composition resulting from the mixing, Jos\'e \& Hernanz (\cite{jh98}) took the composition
of the degenerate cores from Ritossa et al. (\cite{rit96}) for the ONe white dwarfs,
and assumed 49.5\% of oxygen, 49.5\% of carbon and 1\% of neon-22 in the CO cases.
Table \ref{abu} shows the details of the chemical compositions.

\begin{table*}
\centering
\caption{\label{abu} Chemical compositions
of the white dwarf envelope models, in mass fractions. 
Z$_{\rm s}$ contains metals (all elements except H and He) in solar
fraction. $\delta$X$_{i}$ indicates extra mass
fraction of element \emph{i} beyond that included in Z$_{\rm s}$. The total
abundance of metals is indicated by Z$_{\rm total}$.}

\begin{tabular}{ c c c c c c c c c }

\hline\hline
\noalign{\smallskip}
WD & Mixing & Name & X & Y & Z$_{\rm s}$ & & & Z$ _{\rm total}$ \\
\hline 
\noalign{\smallskip}
 & & & & & & $\delta\rm X_{\rm O}$ & $\delta\rm X_{\rm Ne}$  & \\
\hline 
\noalign{\smallskip}
ONe & 75\% & ONe75 & 0.18 & 0.08 & 0.12 & 0.38 & 0.24 & 0.74 \\
ONe & 50\% & ONe50 & 0.35 & 0.15 & 0.09 & 0.25 & 0.16 & 0.5 \\
ONe & 25\% & ONe25 & 0.53 & 0.2 & 0.06 & 0.13 & 0.08 & 0.27 \\
\hline 
\noalign{\smallskip}
 & & & & & & $\delta\rm X_{\rm C}$ & $\delta\rm X_{\rm O}$ & \\
\hline
\noalign{\smallskip}
CO & 50\% & CO50 & 0.35 & 0.14 & 0.02 & 0.245 & 0.245 & 0.51 \\
\hline 

\noalign{\smallskip}
\end{tabular}
\end{table*}


\subsection{Numerical Method and Boundary Conditions}

The numerical code integrates the equations of stellar equilibrium 
from the photosphere to the base of the envelope. 
The envelope is tested at each step for
convective instability according to the Schwarzschild criterion. When
convection appears, the temperature gradient is solved using the standard mixing-length
theory of convection (B\"ohm-Vitense \cite{boh58}).
Radiative opacities have been calculated for each composition with the 
on-line version of the OPAL code from the Lawrence Livermore 
National Laboratory (Iglesias \& Rogers \cite{opal}),
and the energy generation rates for hydrogen
burning are the analytcal approximations 
from Reeves (\cite{reeves}) and Paczy\'nski (\cite{pac83}). The equation
of state is that of an ideal gas with radiation pressure. It has been
checked that for the gas conditions in the present
models, no significant degeneracy (at the base of the envelope)
or partial ionization effects (at the outermost layers) need to be
taken into account, and that the energy production due to helium burning
is negligible.

The stellar equilibrium equations 
are integrated inwards using an embedded (adaptative stepsize controlled)
Runge-Kutta method with Cash-Karp parameters (Cash \& Karp \cite{CashKarp},
Press et al. \cite{NR}).
Integration starts at the photosphere, with a fixed value for
the total luminosity and trial values for the total mass
of the envelope and the photospheric radius, 
and proceeds inwards until the luminosity reaches a sufficiently small value, which is 
considered to indicate the base of the hydrogen burning layer. 
As will be seen later, for all white dwarf masses and
compositions there is a plateau of constant luminosity where it
experiences little change for a wide range of effective temperatures. 
In this plateau, the total luminosity does not determine a unique 
solution anymore and the photospheric radius is fixed instead.

The boundary conditions at the photosphere are obtained from a grey
atmosphere model of the same composition as the envelope, which provides
the value of the atmosphere pressure at the photosphere
for the fixed value of the luminosity and the trial values
of the envelope mass and photospheric radius. 
The atmosphere is tested for convective instability during integration,
which has never appeared before reaching the photosphere in this work.
At the base of the envelope, the envelope mass is required
to be smaller than $10^{-13}\rm M_{\odot}$, while the radius, R$_{\rm base}$,
is compared to the core radius, R$_{\rm c}$,  by
requiring the factor $\left| \frac{\rm R_{\rm base}-\rm R_{\rm c}}{\rm R_{\rm c}}\right|$
to be smaller than $10^{-5}$. If these conditions are not fullfiled,
the integration restarts from the photosphere with new trial values
for the envelope mass and photospheric radius. The core radius is 
taken from the Hamada \& Salpeter (\cite{ham61}) 
core mass-radius relation for zero temperature white dwarfs.
We have checked that a slight increase in the core radius does not affect our main 
results; this radius increase is only expected if degeneracy is 
lifted in the outer core shells as a consequence of heating, which is not probable since 
the timescale for heating propagation towards the core is quite large (Henyey \& 
L'Ecuyer, \cite{hen69}).


\begin{figure*}
\resizebox*{0.99\columnwidth}{!}{\includegraphics{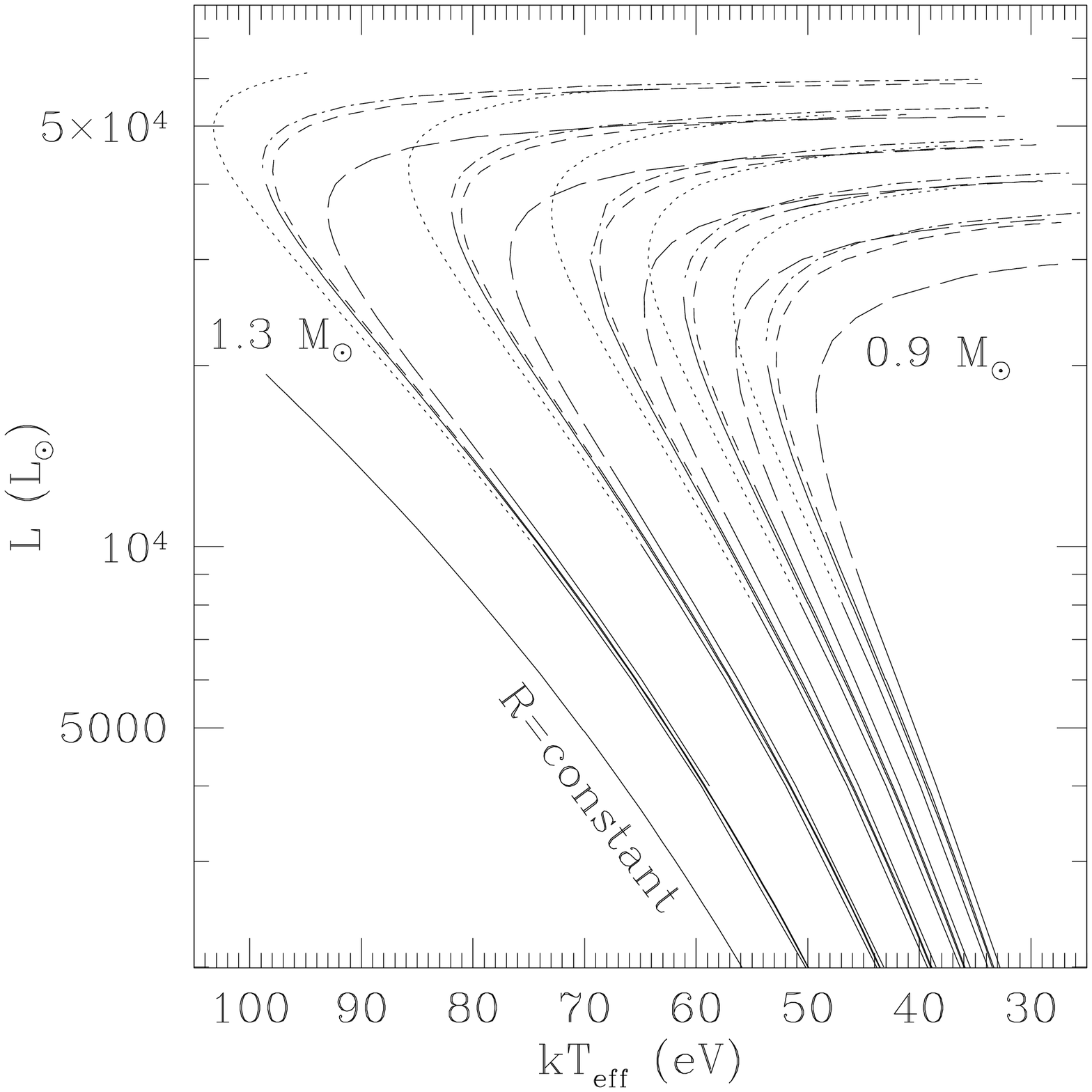}}
\resizebox*{0.99\columnwidth}{!}{\includegraphics{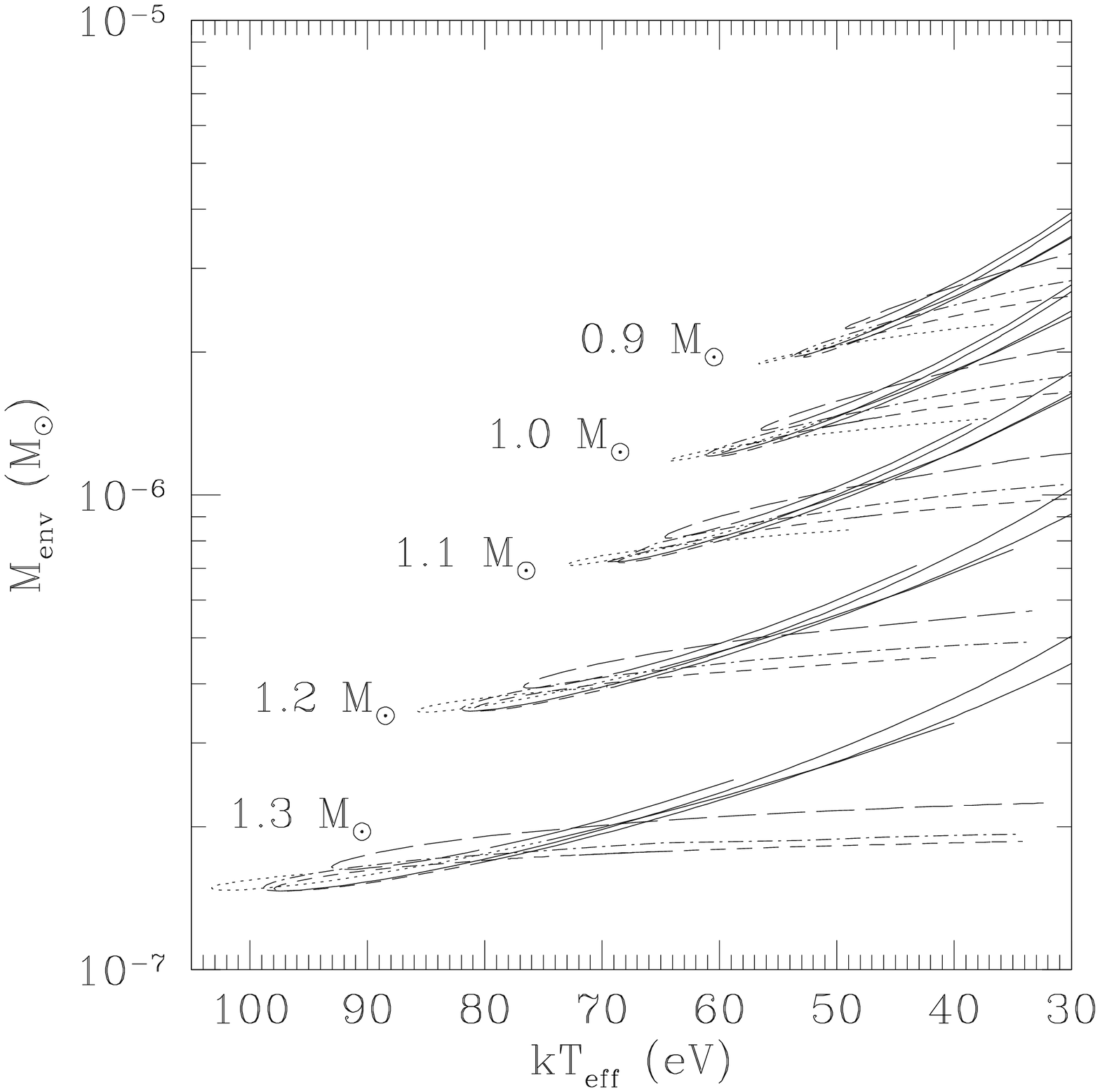}}
\caption{\label{LMkT}Total luminosity (left panel) and envelope mass 
(right panel) versus effective temperature for our white dwarf envelope models:
ONe75 (dotted line), ONe50 (short dash), ONe25 (long dash) and 
CO50 (short dash - dot). Effective temperature is given in eV, 
$kT_{\rm eff} (\rm eV) = 8.617\times10^{-5}T_{\rm eff} (\rm K)$
Solid lines indicate envelopes without convective regions. Five series of
models are plotted for each chemical composition, corresponding to
total masses 0.9, 1.0, 1.1, 1.2  and 1.3 M$_{\odot}$.
A line indicating the luminosity-effective temperature
relation for constant photospheric radius
is over-plotted in left panel for comparison.}
\end{figure*}


\section{Results}

The luminosity versus the effective temperature for all models, 
i.e. their position in the HR diagram, is shown in 
left panel of figure \ref{LMkT}. For each composition and white dwarf mass,
the envelope solutions have a maximum effective temperature that
divides the series of models into two branches: 
a branch of high, quasi-constant luminosity, where the
maximum value for the luminosity is reached; and a branch of low luminosity
and constant photospheric radius. The models of the high luminosity
branch have in all cases a convective region below the photosphere,
while the energy transfer in most of the envelopes of the low
luminosity branch is radiative in the whole structure. 
Right panel in figure \ref{LMkT} shows the 
envelope mass as a function of effective temperature. For
every white dwarf mass there is a minimum value for the envelope
mass, which occurs in the vicinity of the maximum effective temperature.
The envelope mass along the high luminosity branch is almost constant,
decreasing only slowly for increasing effective temperatures, while the 
decrease of the envelope mass for increasing effective temperatures along
the low luminosity branch is steeper. A summary of the main properties of 
all model envelopes is found in table \ref{tabsum}.

\subsection{Stability}

\begin{figure}
\resizebox{0.99\columnwidth}{!}{\includegraphics{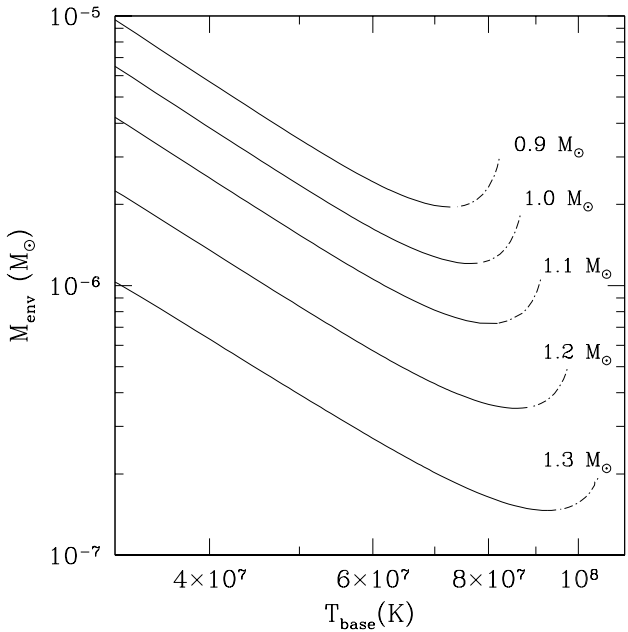}} 
\caption{\label{tmb} Envelope mass versus base temperature for CO50 models.
Solid lines indicate envelopes without convective regions.}
\end{figure}

A stability analysis of the shell source indicates that the burning shell is 
only thermally stable for the solutions along the high luminosity branch. 
The unstability of the low luminosity branch can also be understood in a qualitative way 
if the evolution of the envelope is approximated as a succession of steady 
states with decreasing envelope masses. Figure \ref{tmb} shows the 
temperature at the base of the envelope for CO50 models. The high 
luminosity branch in figure \ref{LMkT} corresponds to envelopes with 
high base temperatures in figure \ref{tmb}, while 
envelopes with low base temperatures correspond to the low luminosity branch.
From left panel in figure \ref{LMkT} it is clear that, along any of the two 
branches, the envelope mass decreases towards higher effective temperatures. 
If the evolution of an envelope is approximated as a 
succession of steady states, it will proceed in 
any case towards decreasing envelope masses, i.e., increasing the effective temperature.
Hydrogen burning will continue
until the minimum envelope mass and maximum effective temperature are reached; 
evolution can not proceed further with stationary hydrogen burning, 
since no equilibrium configuration for a lighter envelope exists.

Along the high luminosity branch, an envelope evolves at quasi-constant luminosity, 
increasing the effective temperature as a consequence of the decrease 
of the photospheric radius, which sinks the photosphere deeper into the envelope.
Nevertheless, the envelope as a whole
is slowly cooling down and the energy production decreases with time,
with both the luminosity and the temperature at the base of the envelope
slowly decreasing (figures \ref{LMkT} -left panel- and \ref{tmb}). 
But along the low luminosity branch 
(the low bottom temperature branch in figure \ref{tmb}), 
for decreasing envelope masses, the temperature at 
the base and the luminosity of the envelope would increase. 
This would require an extra source of energy and indicates that evolution 
cannot proceed along this branch. 

The stable, high luminosity branch is limited at its low effective temperature end by 
the Eddington luminosity. For the low temperature models along this branch, 
the radiative luminosity closely approaches the Eddington limit as effective temperature 
decreases, mainly because there's a photospheric opacity increase leading to an Eddington 
luminosity decrease. Therefore, envelopes in hydrostatic equilibrium do not 
exist for effective temperatures lower, and thus envelope masses and photospheric 
radii larger, than those shown in figure \ref{LMkT}.

\subsection{Interior structure}

\begin{figure*}
\centering
\resizebox*{0.8\columnwidth}{!}{\includegraphics{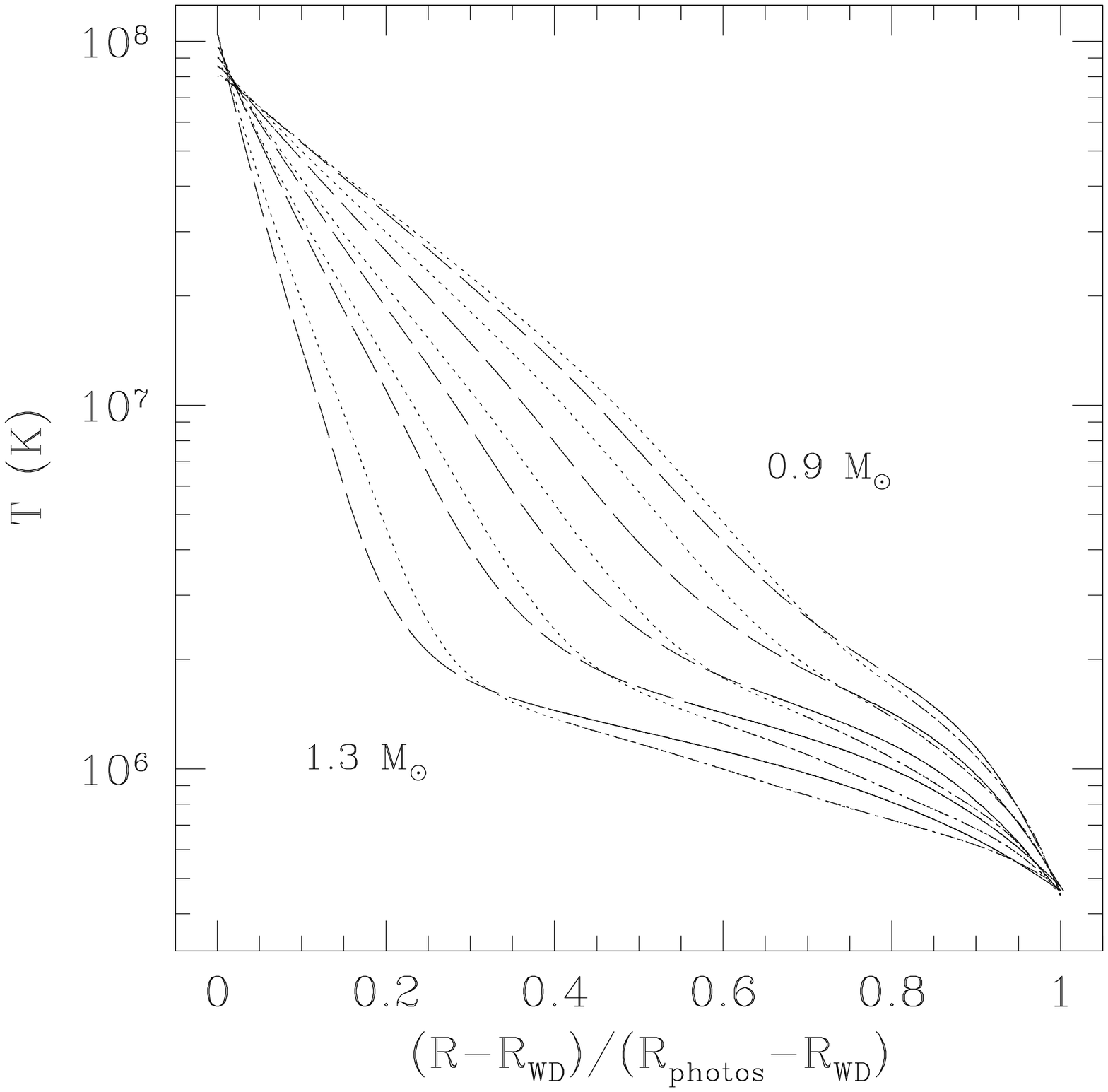}} 
\resizebox*{0.8\columnwidth}{!}{\includegraphics{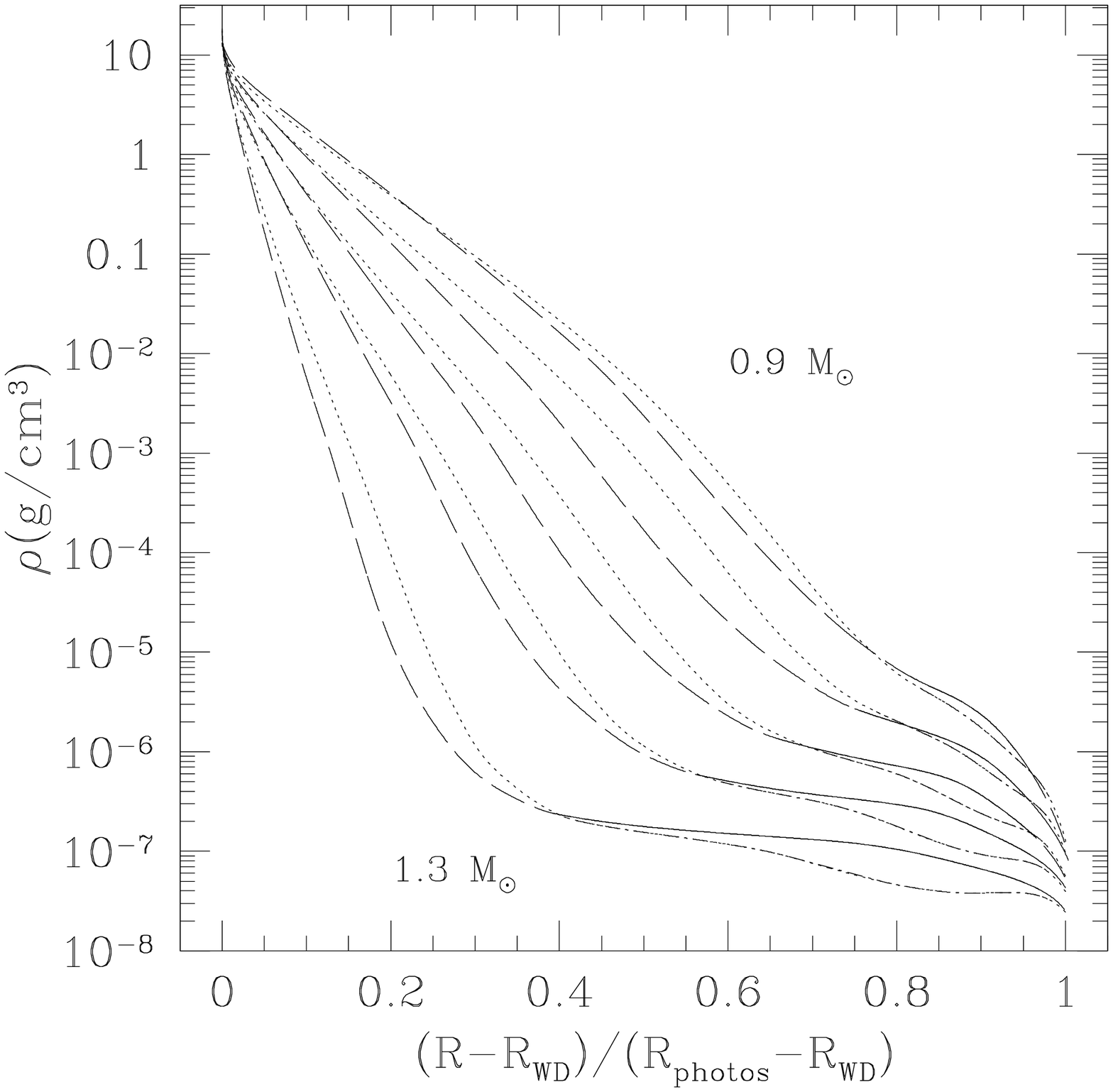}} 

\resizebox*{0.8\columnwidth}{!}{\includegraphics{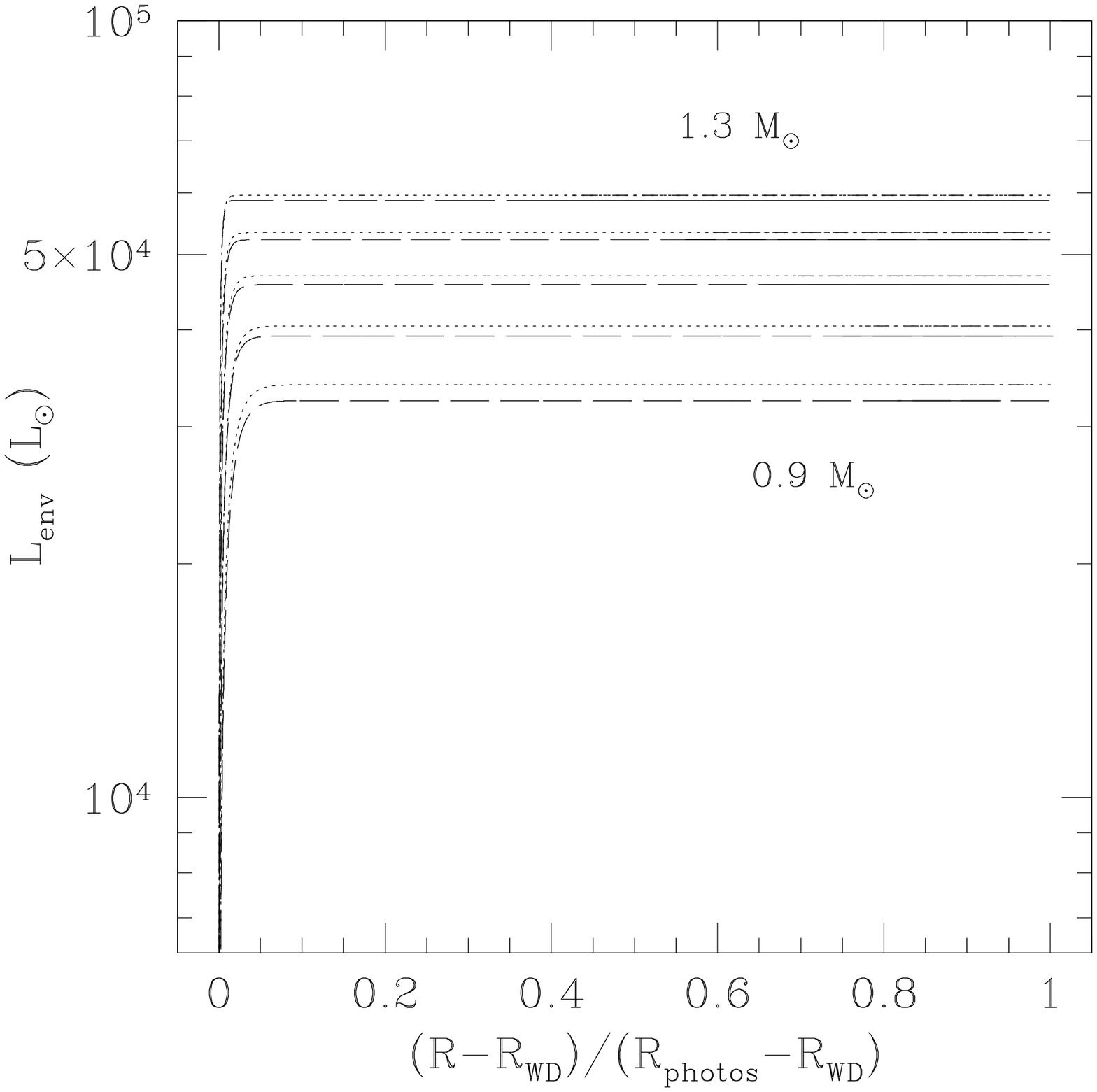}} 
\resizebox*{0.8\columnwidth}{!}{\includegraphics{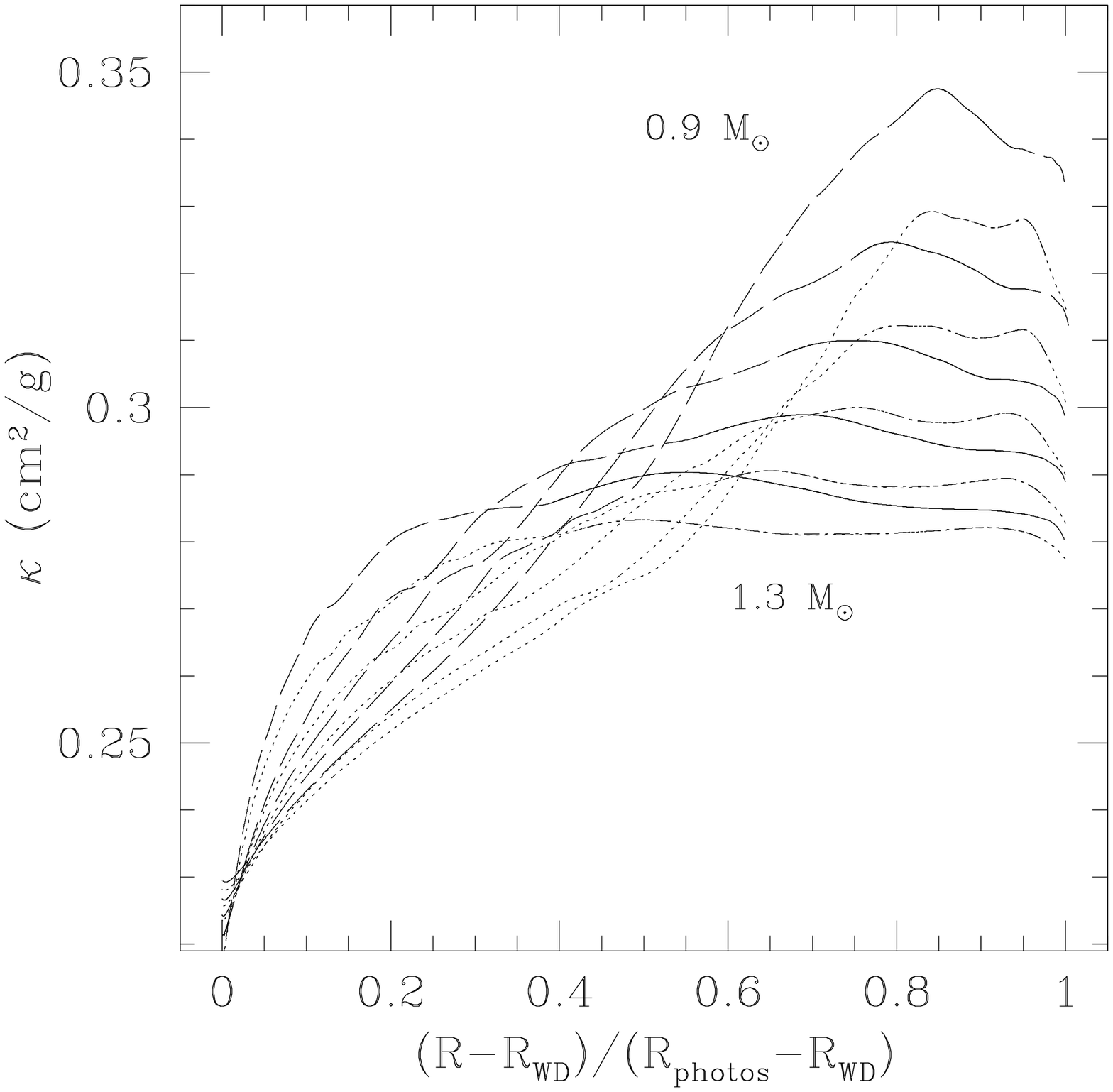}} 

\resizebox*{0.8\columnwidth}{!}{\includegraphics{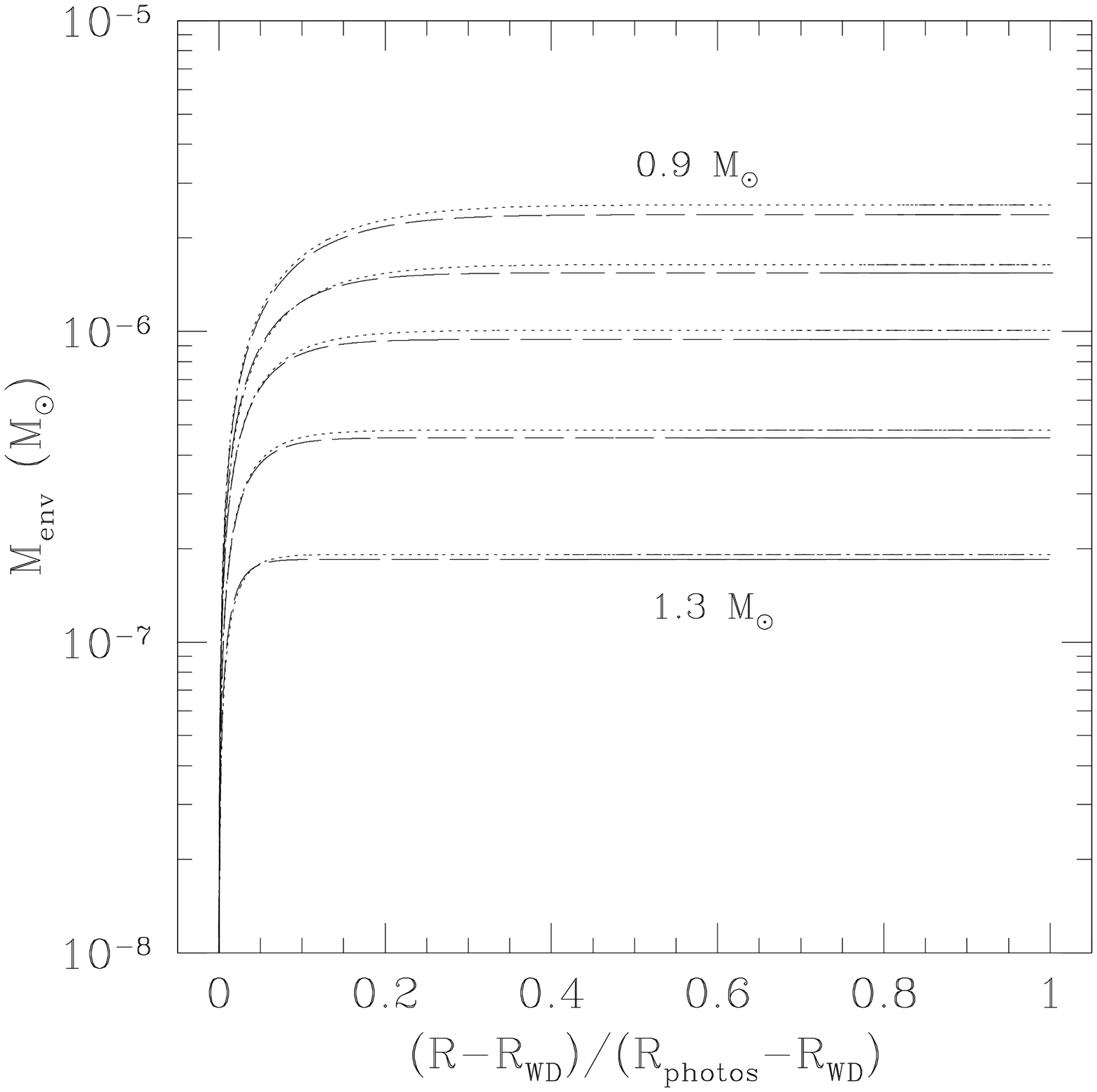}} 
\resizebox*{0.8\columnwidth}{!}{\includegraphics{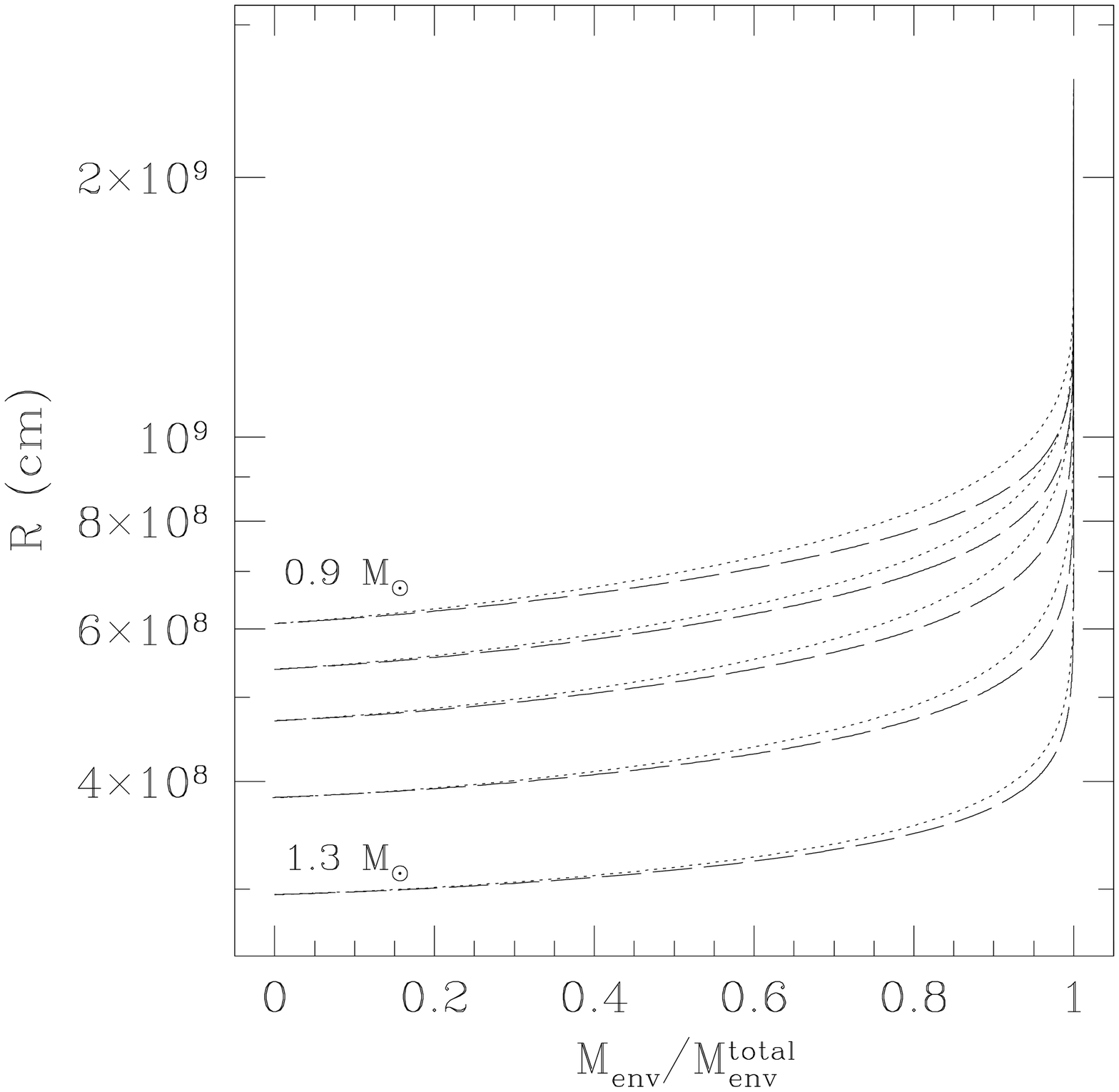}}

\caption{\label{perfils}
Interior structure of ONe50 (dashed line) and CO50 (dotted line) envelopes
with core masses 0.9, 1.0, 1.1, 1.2 and 1.3 M$_{\odot}$,
all of them with a similar effective temperature, $\sim 4.6\times10^{5}$K ($\sim 40$ eV).
Photospheric radii are, for ONe50 models, between 1.9$\times 10^{9}$cm
for 0.9 M$_{\odot}$ and 2.6$\times 10^{9}$cm
for 1.3 M$_{\odot}$; and for CO50 models, between 2.1$\times 10^{9}$cm for 0.9 M$_{\odot}$
and  2.5$\times 10^{9}$cm for 1.3 M$_{\odot}$.
Solid (ONe50) and dash-dotted (CO50) lines indicate convective zones.}

\end{figure*}

Figure \ref{perfils} shows the interior structure of some of the CO50 and ONe50 envelope
models of the high luminosity branch, all of them with the same effective temperature 
($\sim 4.6\times10^{5} \rm K \sim 40$ eV).
The temperature and luminosity profiles show that only the deepest
layers have temperatures high enough to contribute to the energy production.
Luminosity stays constant in most of the envelope. The highest temperatures
at the base are needed for the most luminous envelopes, and in a similar
way the most massive envelopes produce the highest pressures at the bottom.

As mentioned above, envelopes on the high luminosity branch have a
convective region just below the photosphere, where opacity reaches its largest values.
Since convection transfers energy outwards more efficiently than radiation,
the temperature gradient is flatter in the convective regions, as well
as pressure and density gradients, contributing to a larger photospheric
radius and luminosity. 
It can be observed that for the largest luminosities and most
massive white dwarfs, the density gradient in the convective region
is close to zero.
For high luminosities and gravities, inversions
of the density gradient have been frequently found in 
convective layers of white dwarf envelopes
when using the mixing length theory 
(Fontaine \& van Horn \cite{fon76}, Henyey et al. \cite{hen65}, 
Latour \cite{lat70}, Mihalas \cite{mih65}).
However, although the density gradient is very flat in some of the
present models, such inversions do never appear.
The convective region is more extended for increasing white dwarf
masses. As a consequence, temperature, pressure and density gradients
in the inner radiative layers are steeper for massive white dwarfs,
causing the energy production layer to be thinner in these cases,
which is clearly visible in the luminosity profiles. In a similar
way, the envelope mass increases only in the deepest layers in massive
white dwarfs, since most of the envelope has small density values.

\begin{figure*}
\centering 
\resizebox*{0.8\columnwidth}{!}{\includegraphics{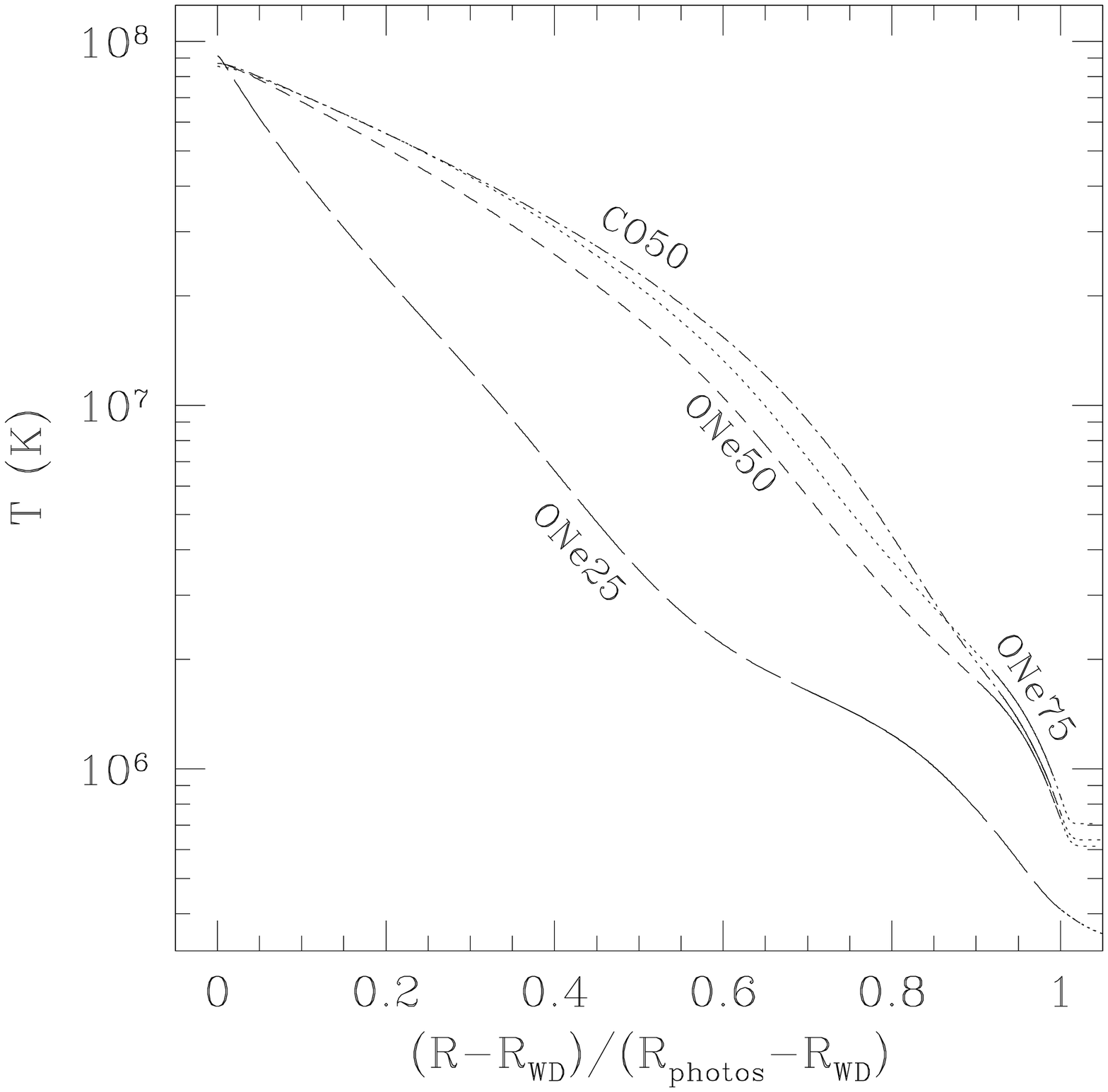}} 
\resizebox*{0.8\columnwidth}{!}{\includegraphics{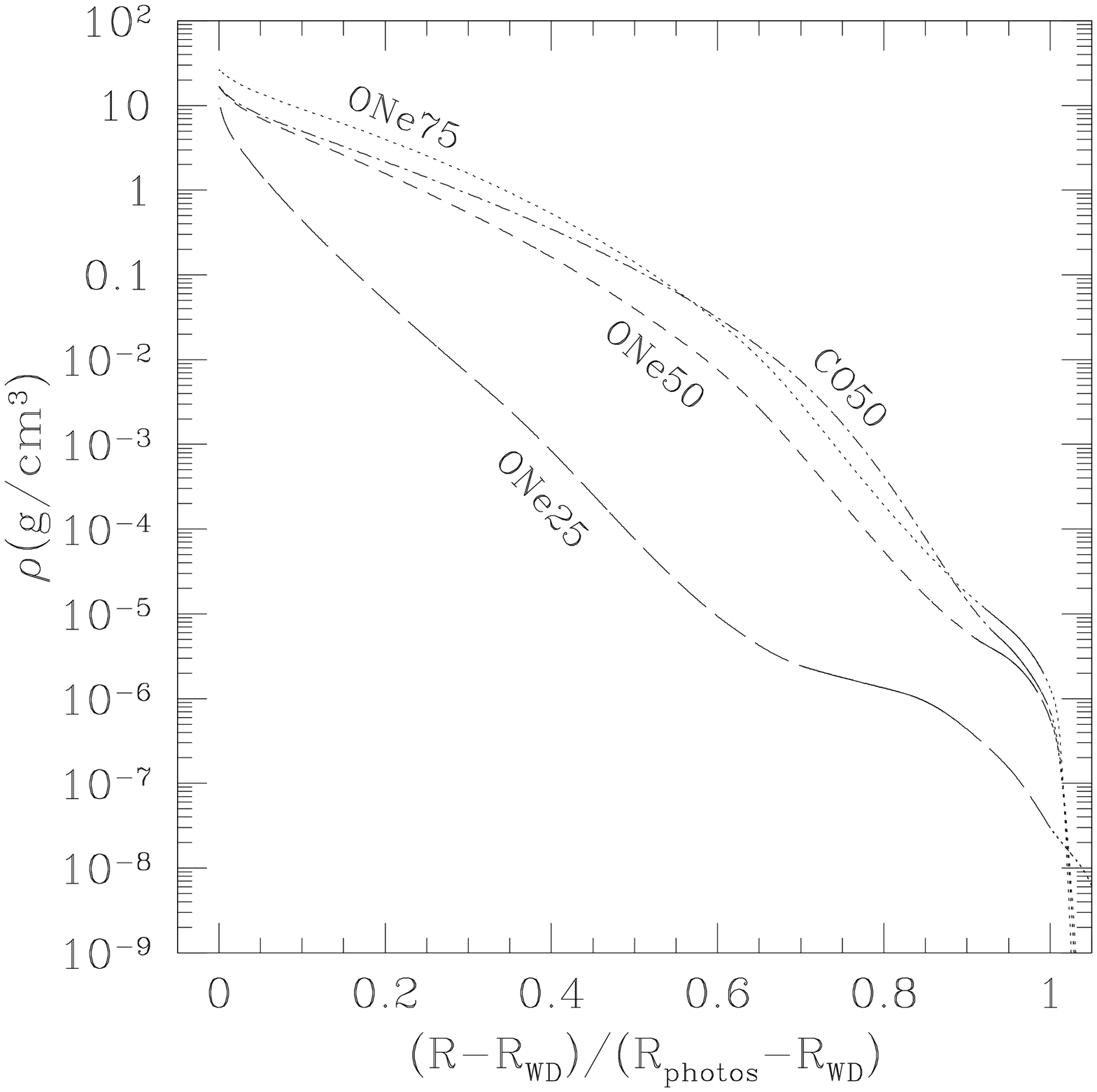}}

\resizebox*{0.8\columnwidth}{!}{\includegraphics{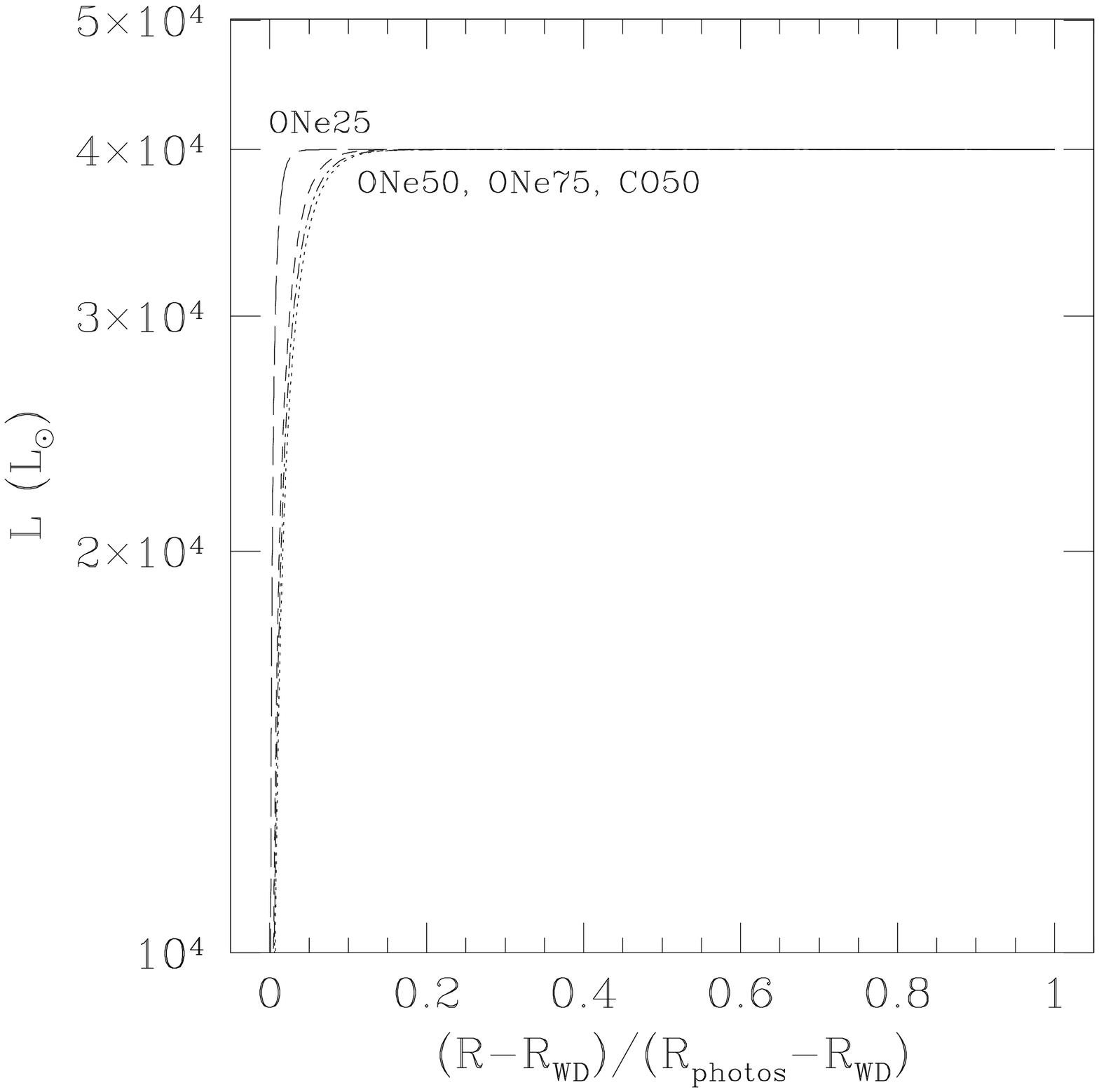}} 
\resizebox*{0.8\columnwidth}{!}{\includegraphics{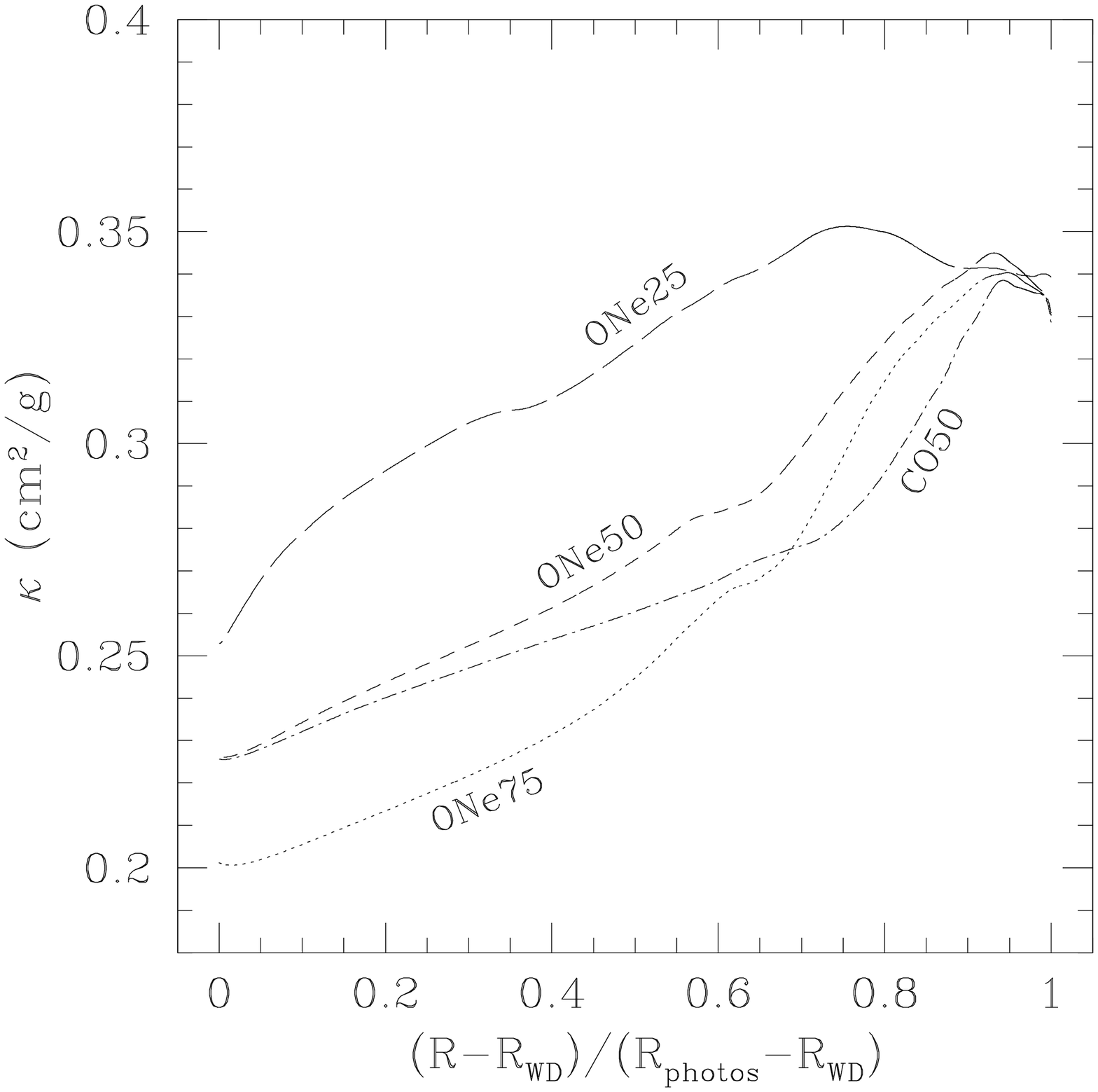}}

\resizebox*{0.8\columnwidth}{!}{\includegraphics{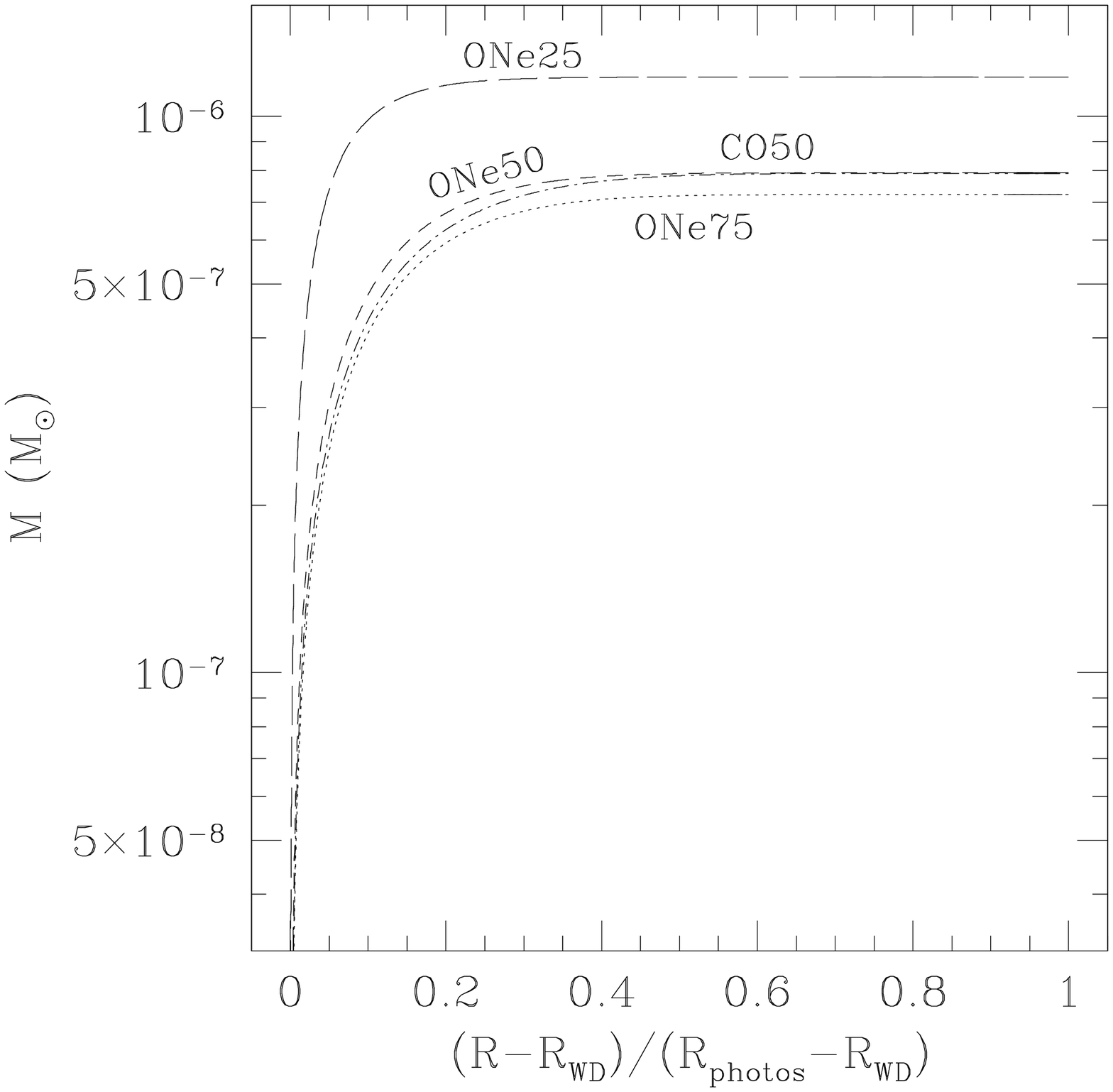}} 
\resizebox*{0.8\columnwidth}{!}{\includegraphics{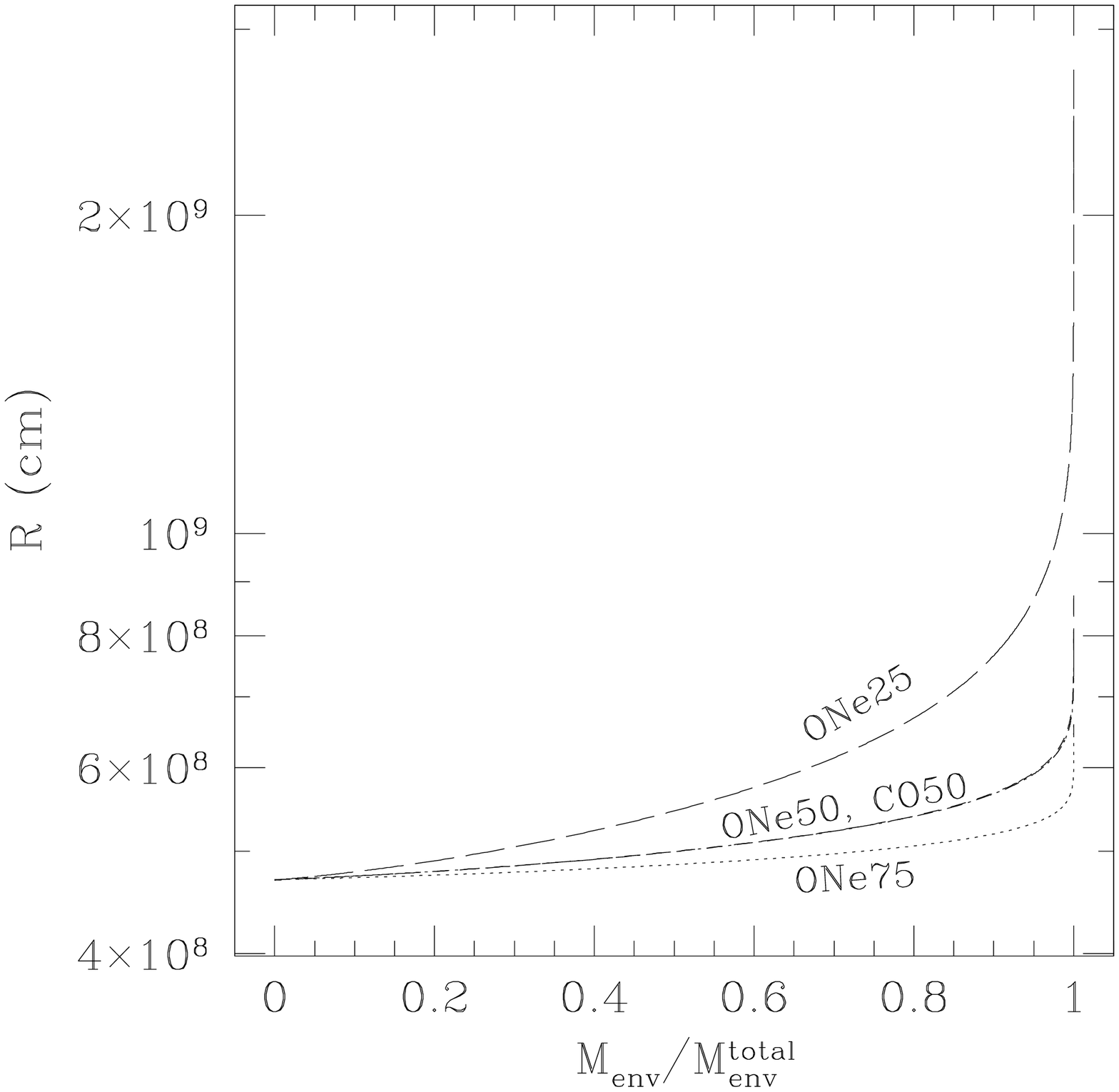}}

\caption{\label{sameMiL}
Interior structure of four envelope models with the same core mass 
(1.1 M$_{\odot}$) and similar luminosity (4$\times 10^{4}\rm L_{\odot}$), 
but with different chemical compositions.  
CO50, with photospheric radius 8.1$\times 10^{8}$cm
(dash dotted line), ONe25, with 2.7$\times 10^{9}$cm
(long-dashed line), ONe50, with 8.7$\times 10^{8}$cm
(short-dashed line), and ONe75, with 6.6$\times 10^{8}$cm
(dotted line). Solid line indicates convection regions, 
and the atmosphere is plotted with dotted line for all cases.}

\end{figure*}

The effects of the different chemical compositions 
on the interior structure are shown in figure \ref{sameMiL},
where profiles for five models with different abundances but the same
core mass and total luminosity are plotted. The strongest difference
in behaviour corresponds to the hydrogen poorest model, ONe75. In
this case, the larger metalicity cause the opacities to be larger
than in hydrogen richer envelopes, causing the convective region to
be much more extended. As a consequence, profiles in the radiative
regions are steeper, specially in the deepest layers.

\begin{figure*}
\centering 
\resizebox*{0.8\columnwidth}{!}{\includegraphics{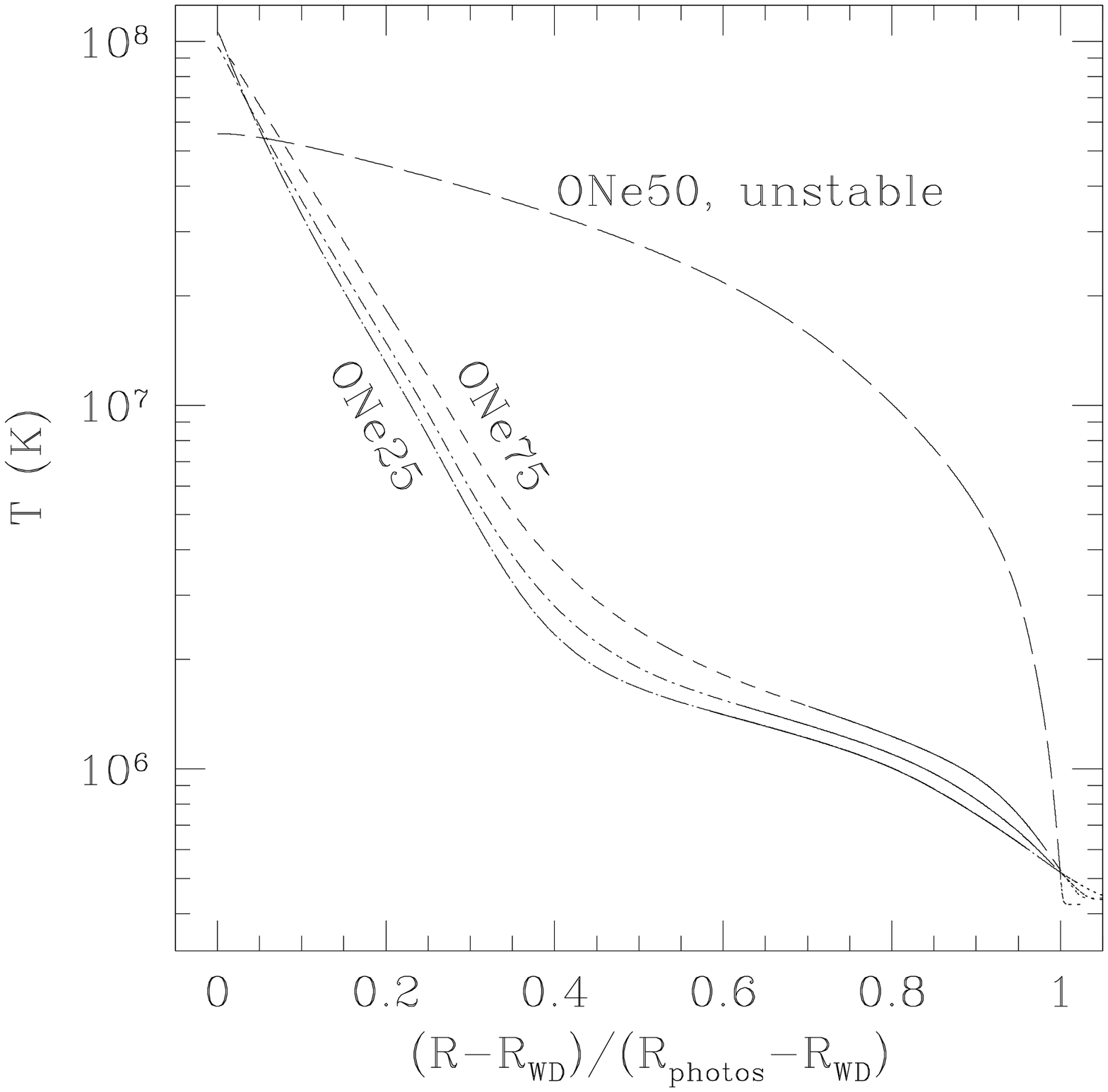}} 
\resizebox*{0.8\columnwidth}{!}{\includegraphics{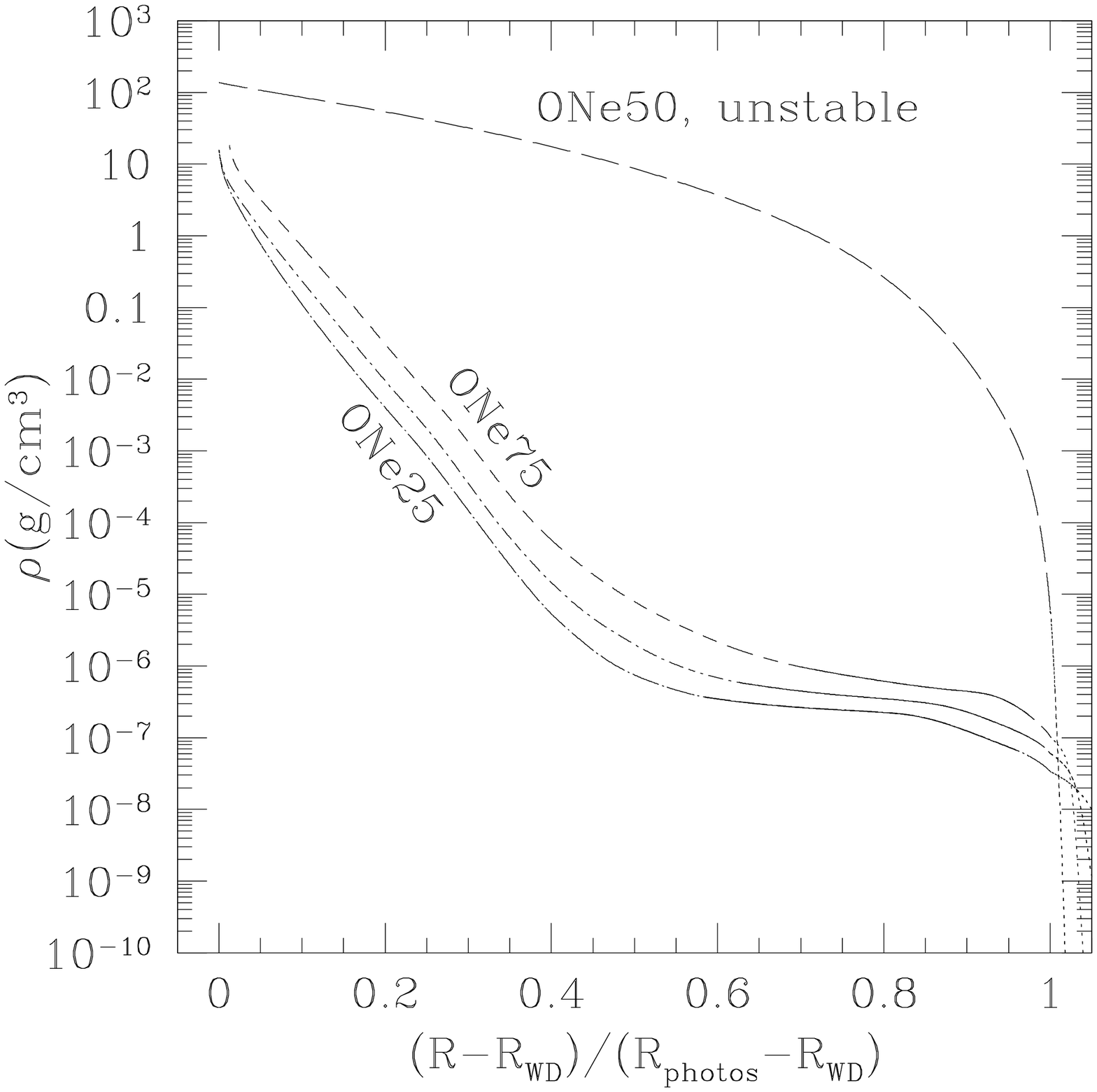}}

\resizebox*{0.8\columnwidth}{!}{\includegraphics{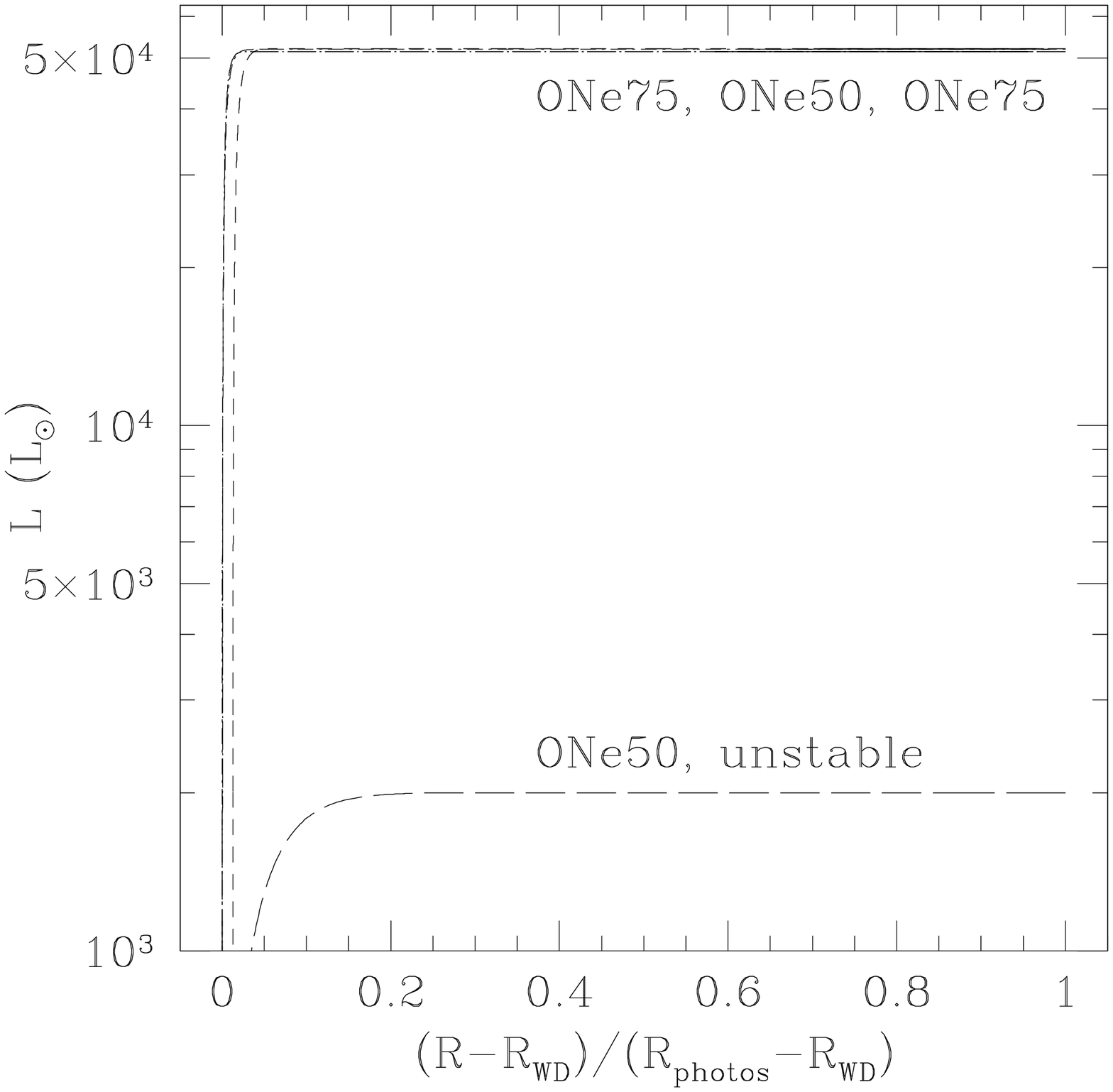}} 
\resizebox*{0.8\columnwidth}{!}{\includegraphics{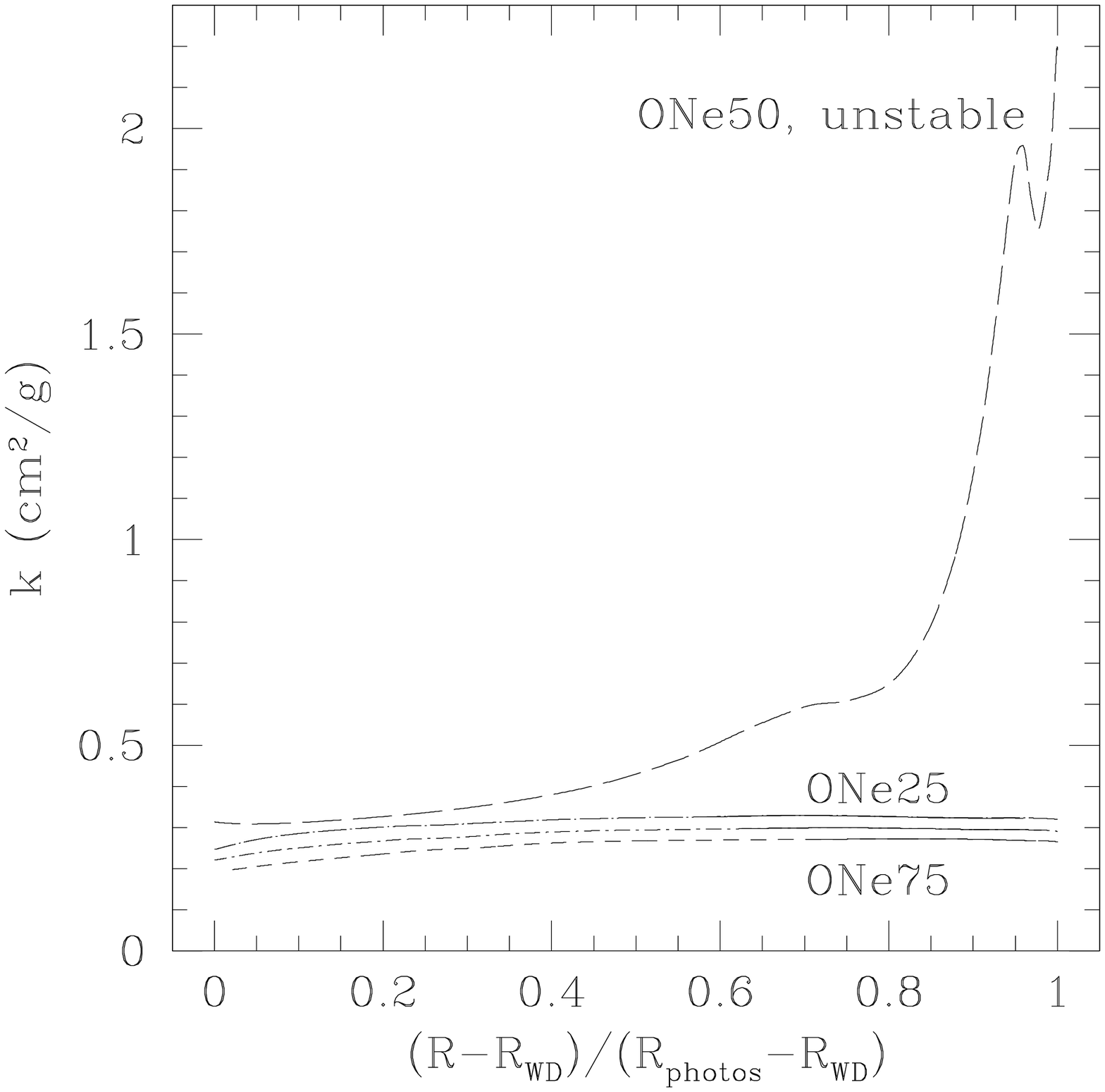}}

\resizebox*{0.8\columnwidth}{!}{\includegraphics{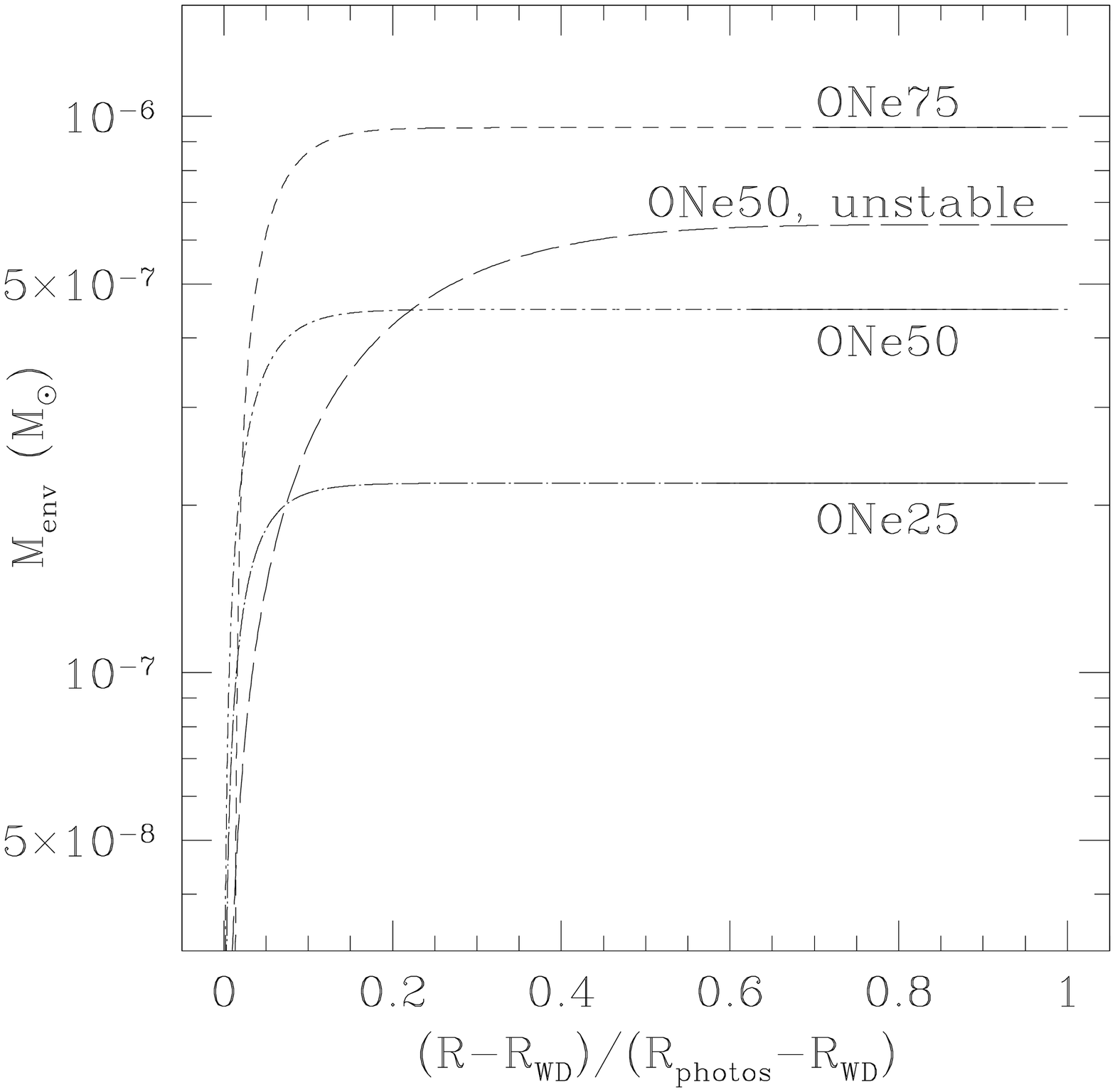}} 
\resizebox*{0.8\columnwidth}{!}{\includegraphics{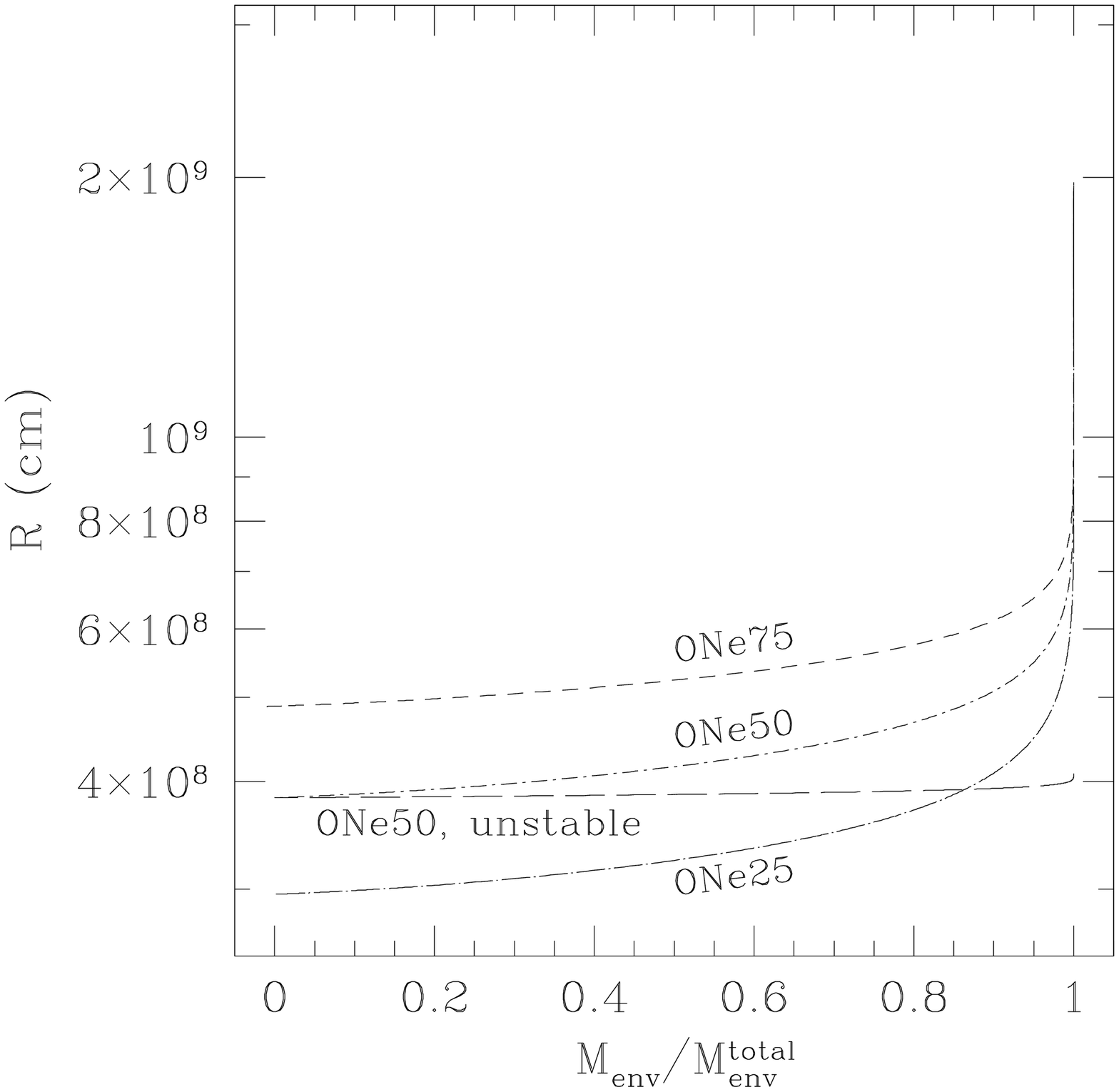}}

\caption{\label{sameLiT}
Interior structure of some ONe white dwarf envelope models with different white
dwarf masses and composition, 
with the same luminosity, $\sim 5.2\times 10^{4}$L$_{\odot}$,
and effective temperature, $\sim 5.2\times 10^{5}$K ($\sim 45$ eV).
Three models belong to the high luminosity branch, all three with
the same total luminosity: ONe25 with M$_{\rm c}=1.3\rm M_{\odot}$ (long dash-dotted line),
ONe50 with M$_{\rm c}=1.2\rm M_{\odot}$ (short dash-dotted line) 
and ONe75 with M$_{\rm c}=1.1\rm M_{\odot}$ (short dashed line). 
All of them have photospheric radius, 2$\times 10^{9}$cm.
A fourth model from the low luminosity branch but with the same effective
temperature is also plotted (ONe50 with M$_{\rm c}=1.2\rm M_{\odot}$,
L=2$\times 10^{3}\rm L_{\odot}$ and R=4$\times 10^{8}$cm,
long-dashed line). Solid lines indicate convective zones, and atmospheres
are plotted with dotted lines.}

\end{figure*}

Different chemical compositions and white dwarf masses combinations
give envelopes with the same observable properties (luminosity and
effective temperature), as can be seen in figure \ref{LMkT}.
Figure \ref{sameLiT} plots the interior structure of three ONe envelopes with the same
effective temperature and luminosity ( $\sim 5.2\times 10^{5}\rm K\sim 45$ eV, 
$\sim 5.2 \times 10^{4}\rm L_{\odot}$),
and thus the same photospheric radius ($2\times 10^{9}$cm),
but different white dwarf masses and compositions (ONe25 with 1.3M$_{\odot}$,
ONe50 with 1.2M$_{\odot}$ and ONe75 with 1.1M$_{\odot}$).
In spite of the differences in abundances, the profiles are very similar,
resulting in equal photospheric properties. 
The convection region is more extended for the hydrogen richest model, ONe25, which is in
general cooler (except for the innermost layers).  
For comparison, an ONe50 model of the unstable, low luminosity branch and with the same effective
temperature is also plotted.

There are small differences in the general properties of
the CO50 and the ONe50 envelope models, despite having the same hydrogen
mass fraction. From figure \ref{LMkT}
it is clear that, for the same effective temperature, the CO50 models
have slightly higher luminosities and larger envelope masses. Equivalently, for the same luminosity,
CO50 envelope models have higher effective temperatures and smaller
photospheric radii. This is evident in figure \ref{perfils},
where CO50 and ONe50 models are plotted for several masses, all of
them with the same effective temperature. For the same effective
temperature, distributions of temperature, pressure and density (in
fraction of total envelope radius) are steeper for ONe50 models. This
is directly related to larger opacities for ONe50 than for CO50 compositions
in the temperature range of the radiative region of the envelopes,
with T$>10^{6}$K, 
that cause the temperature distribution to change more rapidly with radius. Since temperature
decreases more rapidly in the inner regions of ONe50 models than for
CO50, the energy production layer is thinner and the total luminosity
of the envelope is smaller. Analogously, the density profile in the
inner regions is also steeper for ONe50 models, causing the envelope
mass to increase only in a narrower region than in CO50 models and
the total envelope mass to be smaller. 

To check that the differences between CO50 and ONe50 models are due
to the differences in the opacities, some envelope models have been
reintegrated with opacities artificially altered by a certain factor.
We have found that, as in the case of the CO50-ONe50 comparison, envelopes
with higher opacities have smaller luminosities and envelope masses.
A global increase (decrease) of the opacity in the envelope by a 10\%
causes a decrease (increase) in the envelope mass of 4\% and a decrease
(increase) in the luminosity of 10\%.

\section{Analytical relations between main envelope properties}

\subsection{Core mass - luminosity relation}

\begin{figure*}
\resizebox*{0.99\columnwidth}{!}{\includegraphics{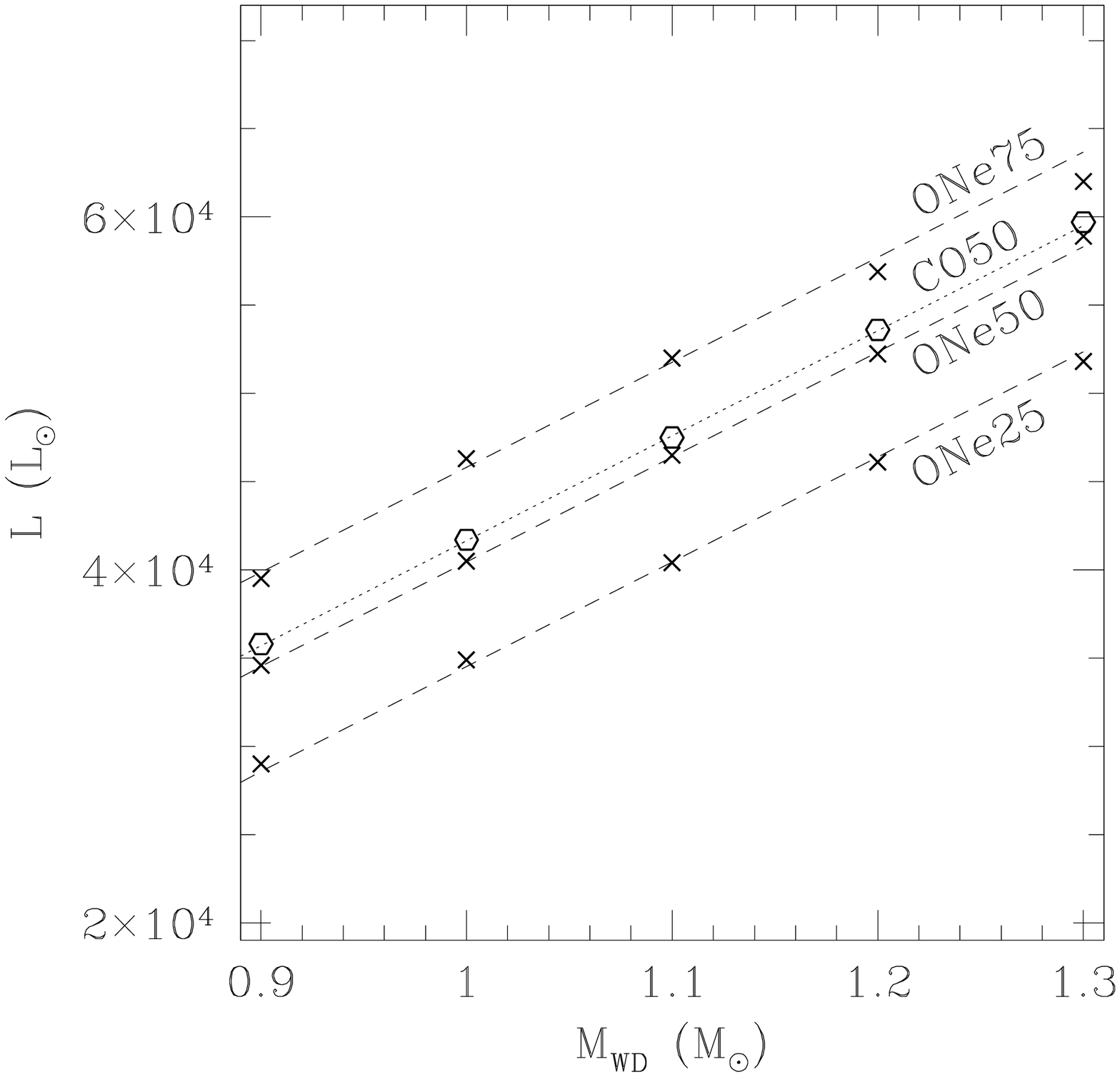}} 
\resizebox*{0.99\columnwidth}{!}{\includegraphics{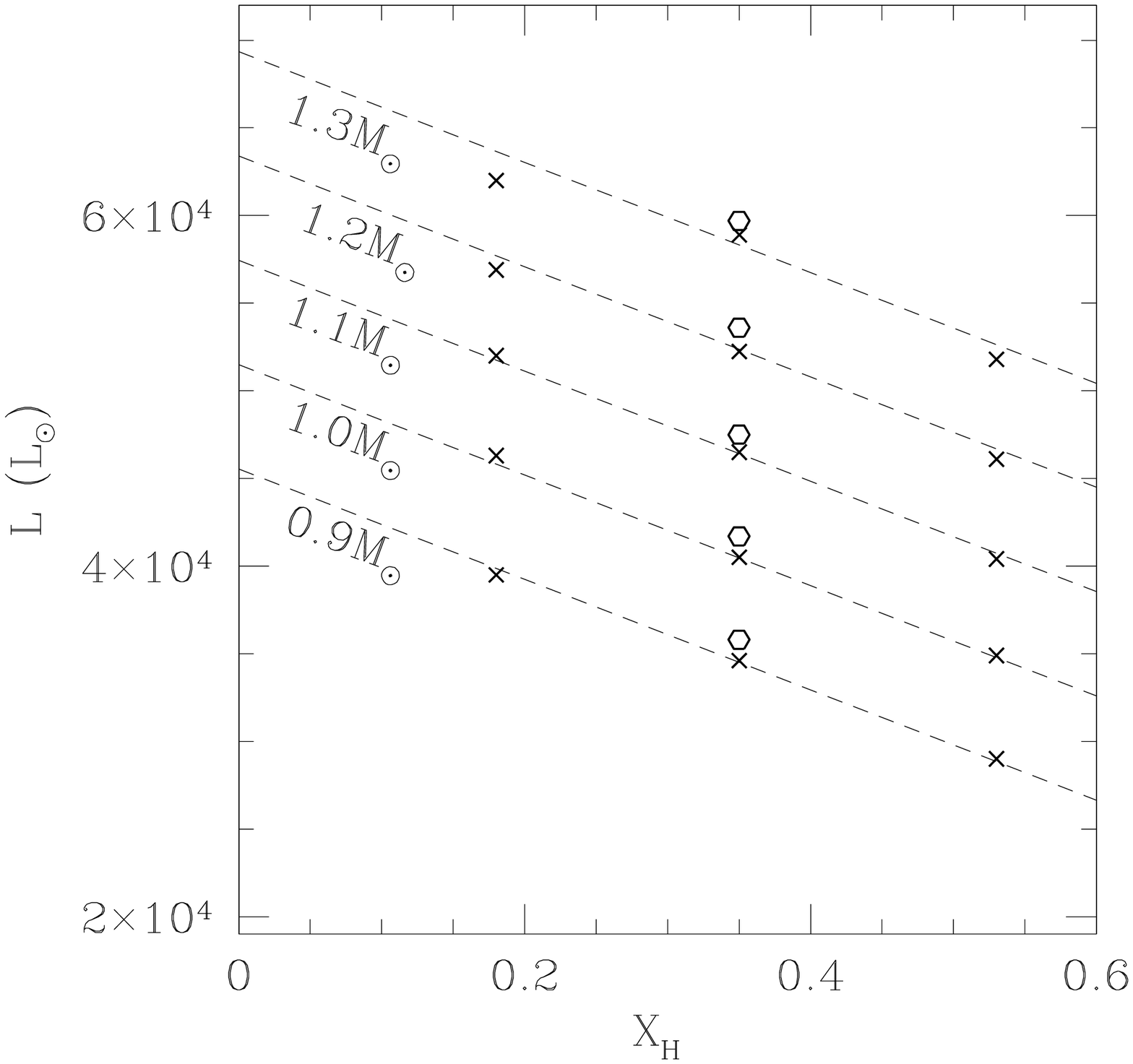}}
\caption{\label{CMLR}
Plateau luminosity versus core mass (left) and hydrogen mass fraction (right).
Crosses indicate ONe envelope models results, while open dots correspond to CO50 models.
Dashed lines correspond to relation \ref{equXLR} for ONe models, and 
the dotted line in the left panel to relation \ref{equCMLRCO50} for CO50 models. 
}
\end{figure*}

From figure \ref{LMkT} it is clear
that luminosity is almost constant along the stable, high-luminosity
branch, and that its value increases with the core mass. In the left
panel of figure \ref{CMLR},
plateau luminosities are plotted against core mass for all models.
The resulting core mass - luminosity relation for our ONe models 
can be approximated by
\begin{equation}
\label{equXLR}
L_{\rm ONe}^{\rm plateau}(L_{\odot })\simeq 5.95\times 10^{4}\left( \frac{M_{\rm c}}{M_{\odot }}-0.536X_{\rm H}-0.14\right) 
\end{equation}
and for the CO50 models by 
\begin{equation}
\label{equCMLRCO50}
L_{\rm CO50}^{\rm plateau}(L_{\odot })\simeq 5.95\times 10^{4}\left( \frac{M_{\rm c}}{M_{\odot}}-0.3\right) 
\end{equation}
The plateau luminosity increases with the white dwarf mass, 
and for increasing mixing factors of the accreted solar matter with
the core material (i.e., for decreasing hydrogen abundances).
Envelopes with a lower hydrogen mass fraction need higher temperatures
to reach an equilibrium configuration, and therefore have larger luminosities.
The plateau luminosity versus the hydrogen mass 
fraction is shown in the right panel in figure \ref{CMLR}
for all models. Relation (\ref{equXLR}) is over-plotted
for each core mass. While ONe models follow (\ref{equXLR})
(except for the most massive white dwarf),  CO50 models follow a parallel relation but 
are slightly more luminous than ONe50 (between a 3\% and a 7\% more luminous), 
which can also be seen in figure \ref{LMkT}.

The linear relation between the luminosity of a shell burning star on a degenerate core 
and its core mass has been known since long.
An analytical justification for this linear dependence
was obtained by Kippenhahn (\cite{kip81}) (see also
Kippenhahn \& Weigert \cite{kip90}). 
He showed that the luminosity was directly proportional to the core mass and 
independent of its radius for $\beta \ll 1$, where $\beta = \frac{\rm{P}_{\rm{gas}}}{\rm{P}_{\rm{total}}}$.
With this analytical work, Kippenhahn (\cite{kip81}) confirmed the results
found with numerical evolutionary calculations by Paczy\'nski (\cite{pac70}),
who showed that the luminosity of post-AGB stars with a double burning
shell (H, He) on top of a mostly degenerate CO core could be expressed
as $L(L_{\odot})=a(M_{\rm c}/M_{\odot}-b$),
with \emph{a} = 59,250 and \emph{b} = 0.522. Later works found similar
results, with small changes in the parameters. Among them, Iben (\cite{ibe82}),
found that in the case of steady state, \emph{a} = 59,500 and \emph{b} = 0.5. 
For the hydrogen mass fraction used by Iben (\cite{ibe82}) (X$_{\rm H}$=0.64),
the 0.536 X$_{\rm H}$+0.14 factor in equation (\ref{equXLR}),
0.48, is very similar to the value found in that work, 0.5.

The increase of the luminosity with decreasing hydrogen mass fraction
was also found in analytical studies by Tuchman et al. (\cite{tuc83})
and with numerical calculations of steady hydrogen burning on white
dwarfs by Tuchman \& Truran (\cite{tuc98}). They studied the dependence of 
the luminosity with the hydrogen mass fraction 
for a fixed metalicity (Z=0.25), 
and for a constant He/H ratio (assumed to be solar, $\approx$ 0.1), 
and summarized their dependences by the relation 
L(L$_{\odot}$)=5.2$\times 10^{4}[$M$_{c}/$M$_{\odot}$-0.205-0.5(X-Z)].
According to the authors, this relation holds also for other metalicities.
More recently, Marigo (\cite{mar00}) used an homology formalism 
to justify the linear core mass - luminosity relation
and study the dependence of the luminosity with the chemical composition, 
for giant stars with core masses between 0.5 and 0.7 M$_{\odot}$.

\begin{figure*}
\resizebox*{0.99\columnwidth}{!}{\includegraphics{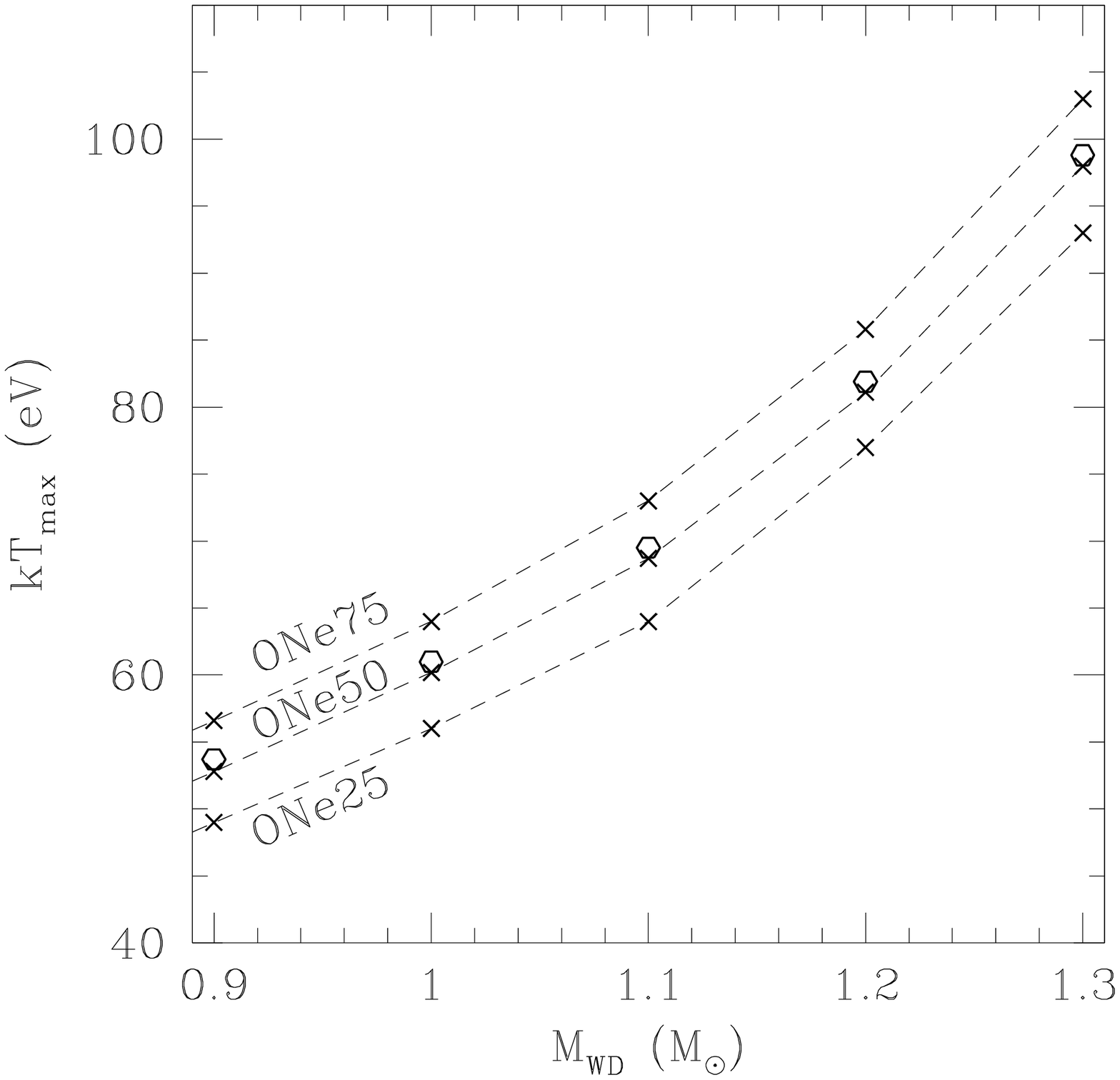}} 
\resizebox*{0.99\columnwidth}{!}{\includegraphics{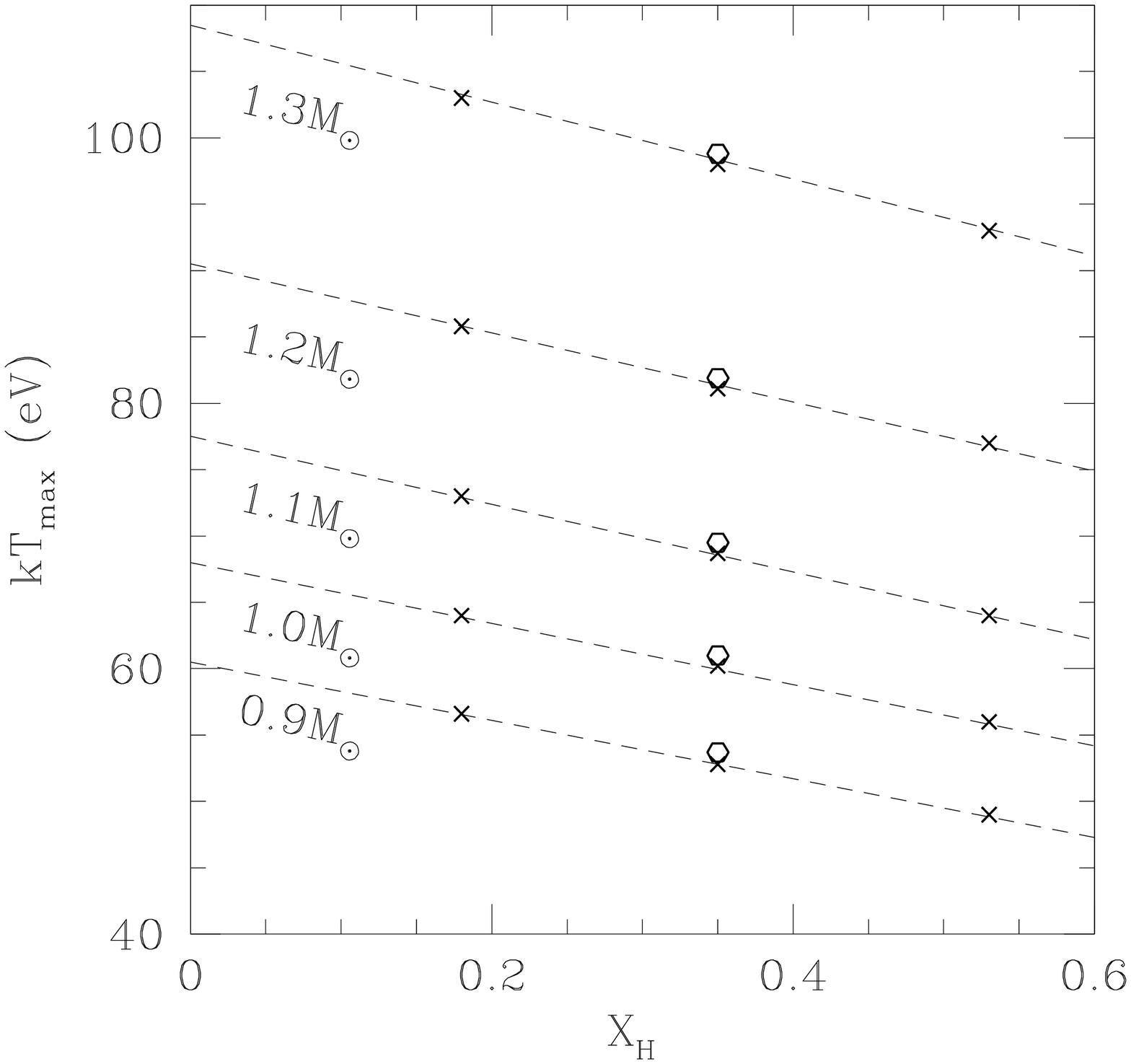}}
\caption{\label{kTMX}
Maximum effective temperature versus core mass (left) and 
hydrogen mass fraction (right) for the same models shown in figure \ref{CMLR}. 
Lines joining points are for orientation purposes.}
\end{figure*}

\begin{figure*}
\resizebox*{0.99\columnwidth}{!}{\includegraphics{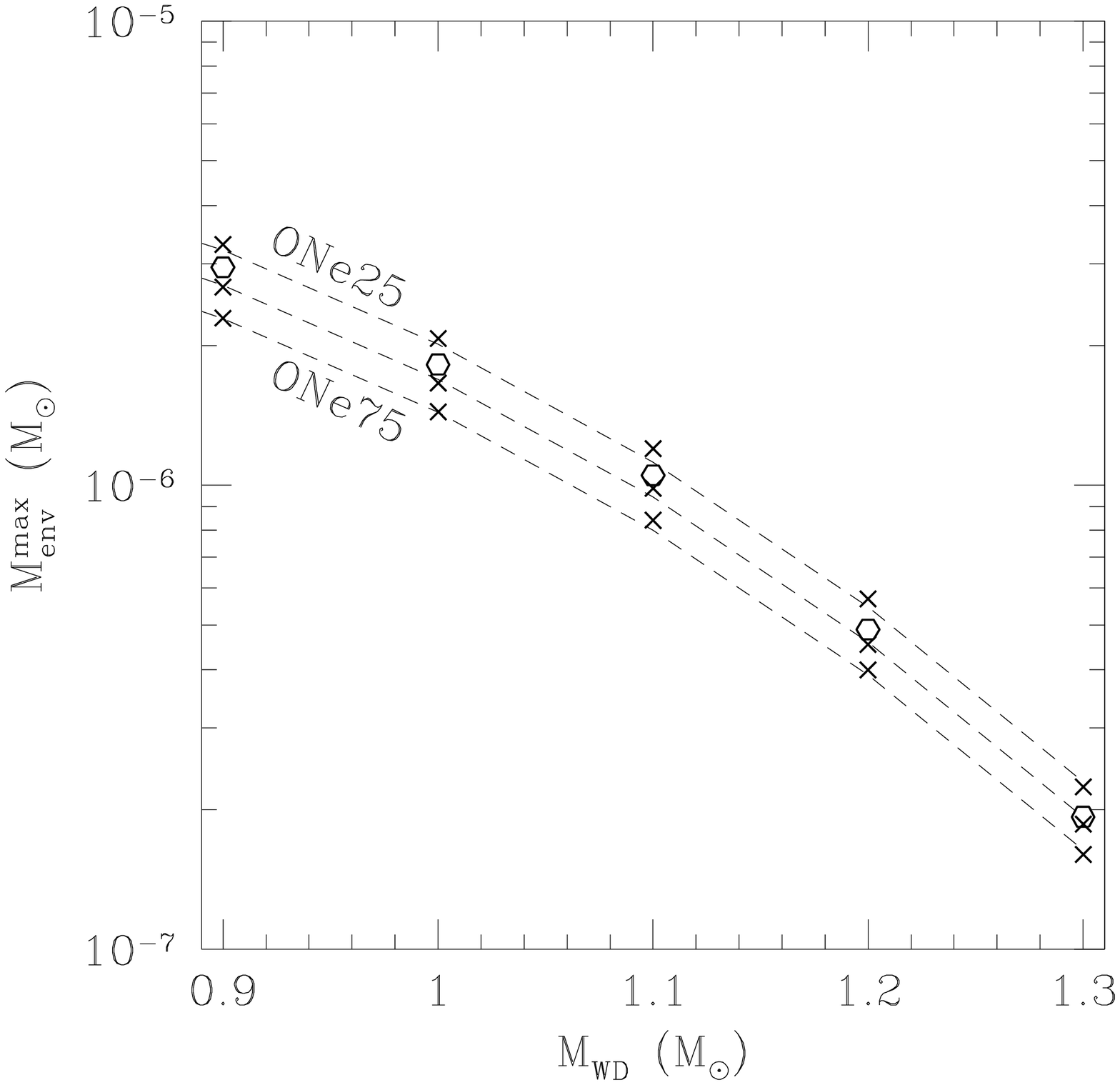}} 
\resizebox*{0.99\columnwidth}{!}{\includegraphics{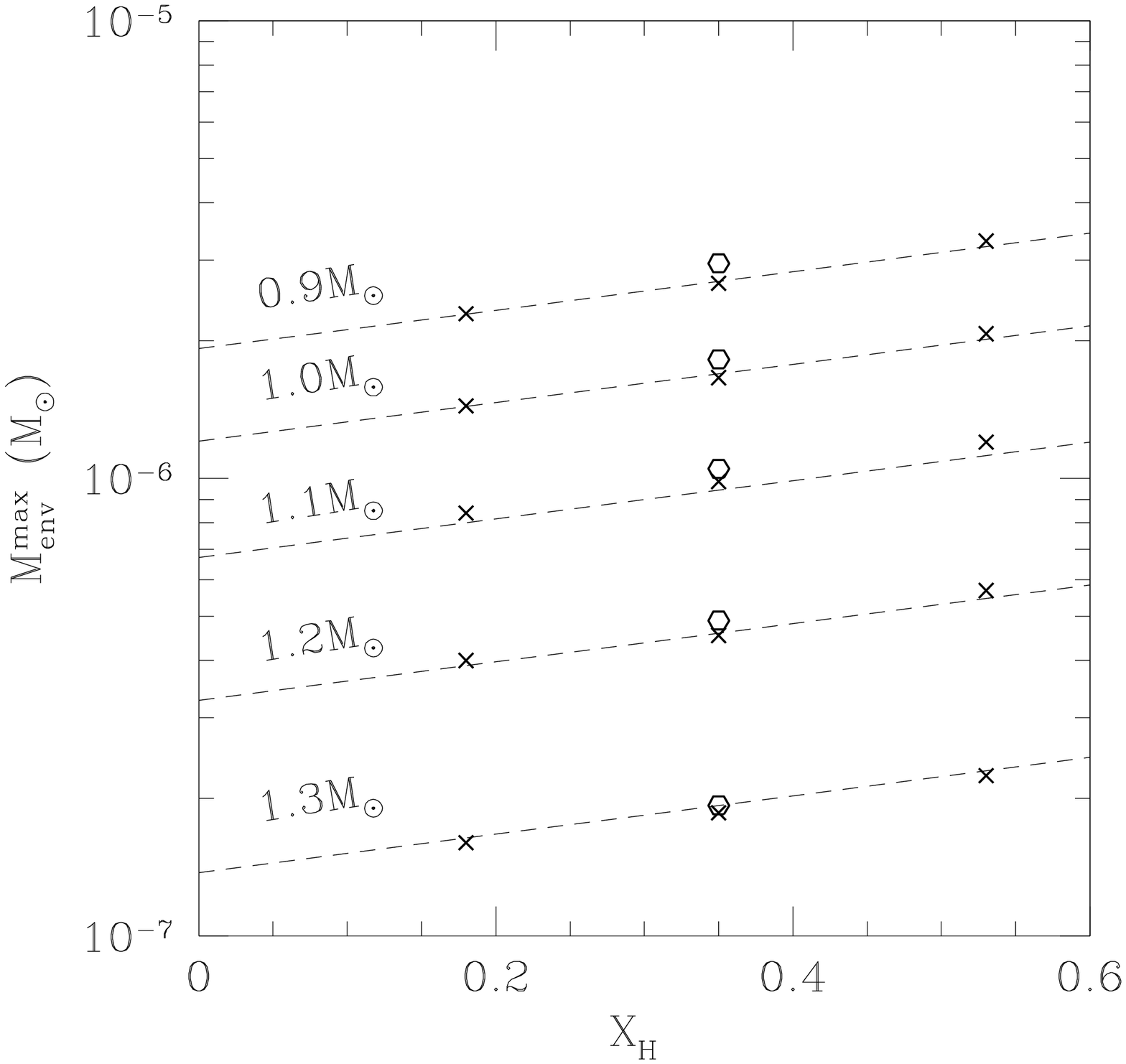}}
\caption{\label{MMX}
 Maximum envelope mass versus core mass (left) and hydrogen 
mass fraction (right) for the same models shown in figure \ref{CMLR}. 
Relation (\ref{equMMX}) is shown in dashed lines.}
\end{figure*}

\begin{figure*}
\resizebox*{0.99\columnwidth}{!}{\includegraphics{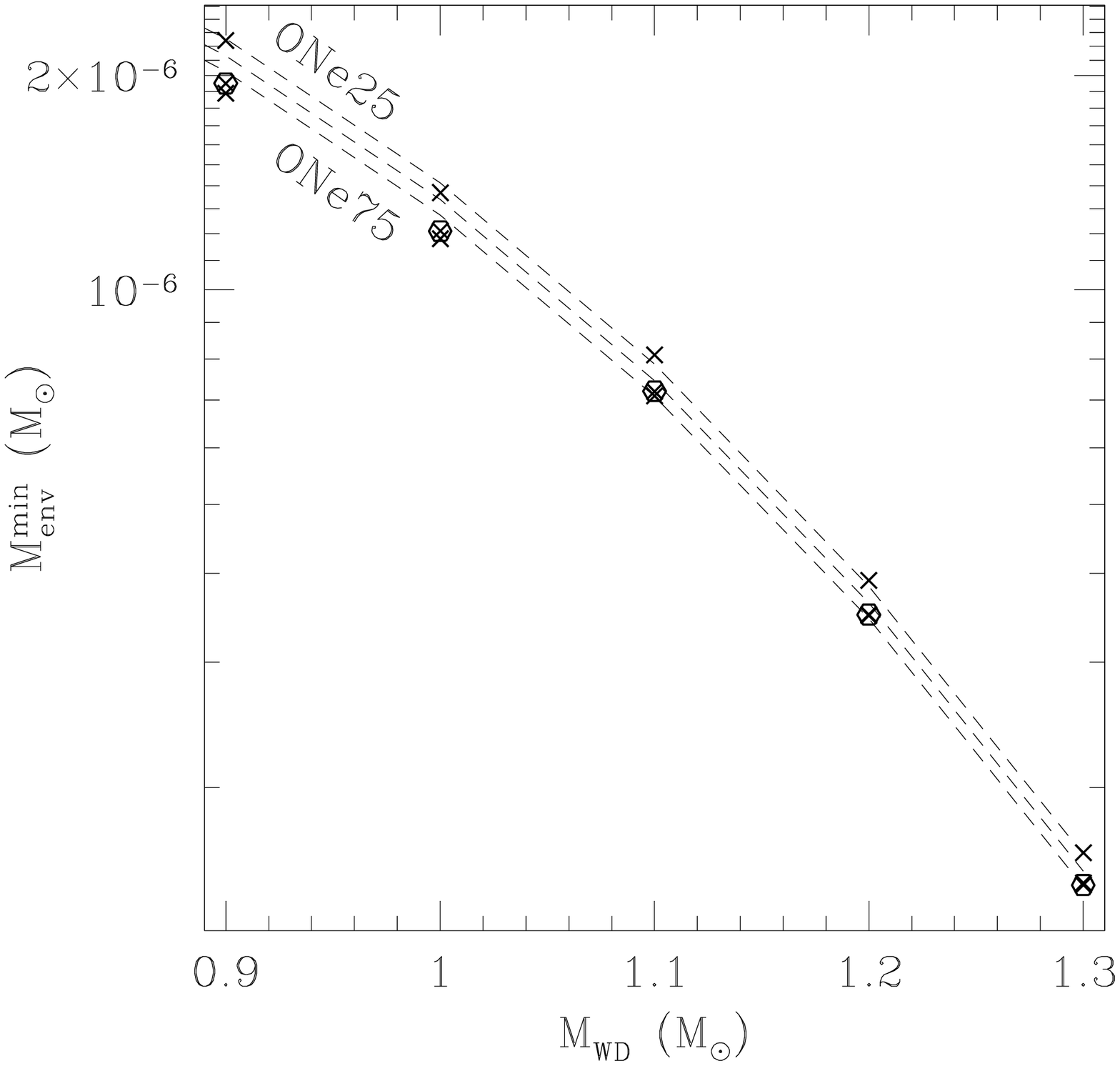}} 
\resizebox*{0.99\columnwidth}{!}{\includegraphics{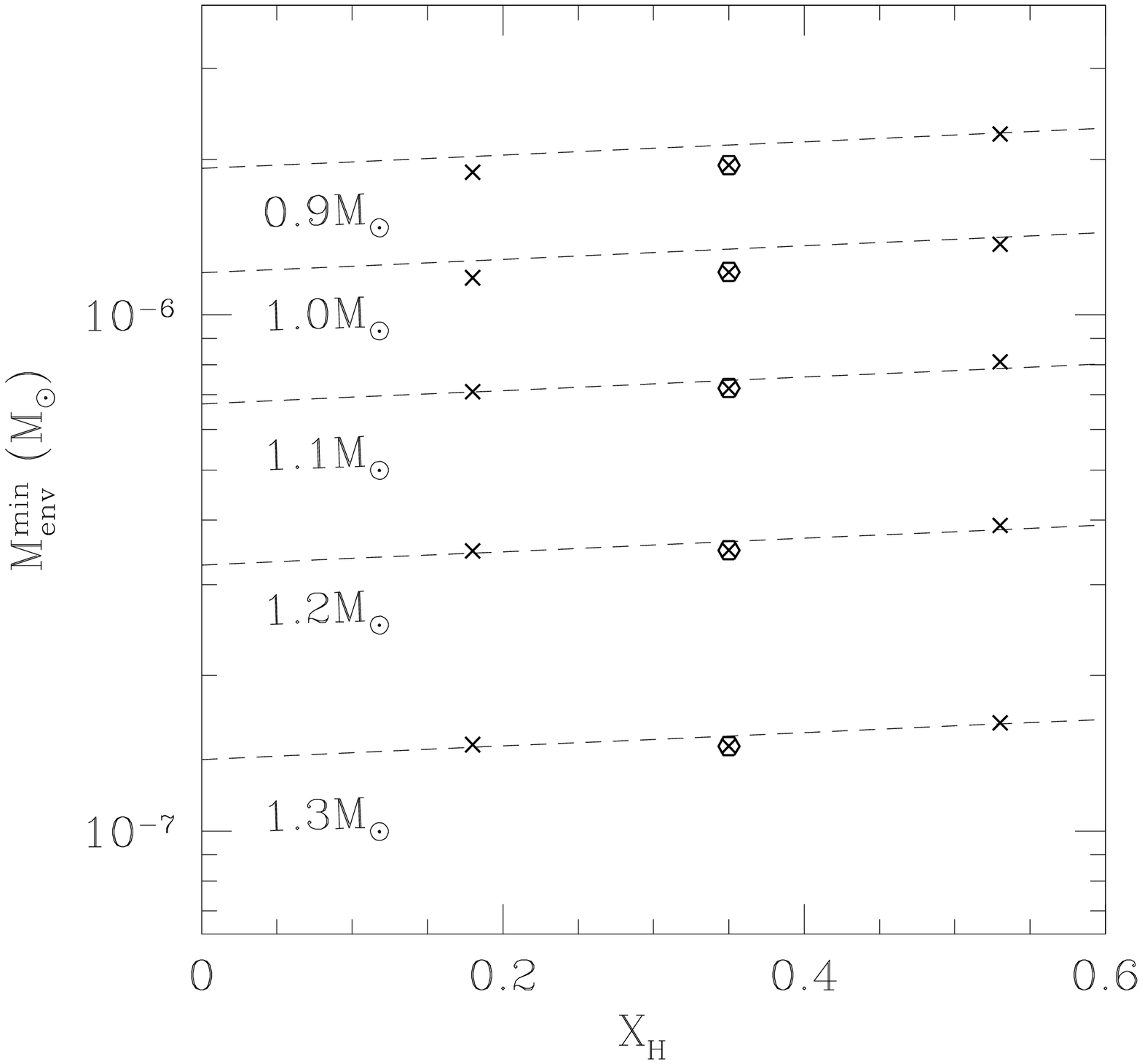}}
\caption{\label{Mmin}
Minimum envelope mass versus total mass (left) and hydrogen mass fraction (right) 
for the same models shown in figure \ref{CMLR}. The minimum envelope mass
depends only slightly on the hydrogen mass fraction, resulting in very similar values for models
with different composition (overplotted points for ONe75, ONe50 and CO50 models in left panel).
Dashed lines correspond to relation (\ref{equMmin}).}
\end{figure*}


\subsection{Other relations}

The maximum effective temperature also increases with decreasing hydrogen
mass fractions and with increasing core masses (see figure \ref{kTMX}),
but in this case, no clear relation has been found. As shown in the
figure, the maximum effective temperature does not follow any unique linear
relation or power law with the core mass. 
Its behaviour can be only approximated by a power law, 
$kT_{\rm eff}(\rm eV)\propto M_{\rm c}^{9/10}$ (for core
masses smaller than 1.1M$_{\odot}$) and $kT_{\rm eff}(\rm eV)\propto M_{\rm c}^{5/9}$ 
(for more massive white dwarfs).

As in the case of the luminosity, the envelope mass in the luminosity plateau depends on
the core mass and on the hydrogen mass fraction (see figure \ref{LMkT}, right panel). 
Iben (\cite{ibe82}) found that the relationship between the envelope mass and the white dwarf
mass in the luminosity plateau was 
$log M^{\rm Iben82}_{\rm env}(M_{\odot})\simeq -4.38-4.96 (M_{\rm c}/M_{\odot}-1)$. 
Since the envelope mass decreases along the luminosity plateau for increasing 
effective temperatures (see figure \ref{LMkT}) a clear definition of the envelope 
mass can not be established. In figure \ref{MMX} (left panel), the maximum envelope 
mass for our stable models is plotted as a function of the core mass. It is clear that
for our models, the $log M^{\rm max}_{\rm env}-M_{\rm c}$ relation is not linear. 
A general relation between the maximum envelope mass, the core mass and the hydrogen mass 
fraction can be found for our envelopes:

\begin{equation}
\label{equMMX}
\log {M^{\rm max}_{\rm env}}(M_{\odot})\simeq 0.42 X_{\rm H}-\left( \frac{M_{\rm c}}{M_{\odot }}-0.13\right) ^{3}-5.26
\end{equation}

\noindent
which is indicated in figure \ref{MMX}.

The minimum envelope mass does not have a strong dependence on the
hydrogen mass fraction (see right panel of figure \ref{MMX}),
and can be approximated by the relation
\begin{equation}
\label{equMmin}
\log M^{\rm min}_{\rm env}(M_{\odot })\simeq 0.13X_{\rm H}-\left( \frac{M_{\rm c}}{M_{\odot }}-0.13\right) ^{3}-5.26
\end{equation}

\section{Quasi-static temporal evolution}

\begin{figure*}
\resizebox*{0.99\columnwidth}{!}{\includegraphics{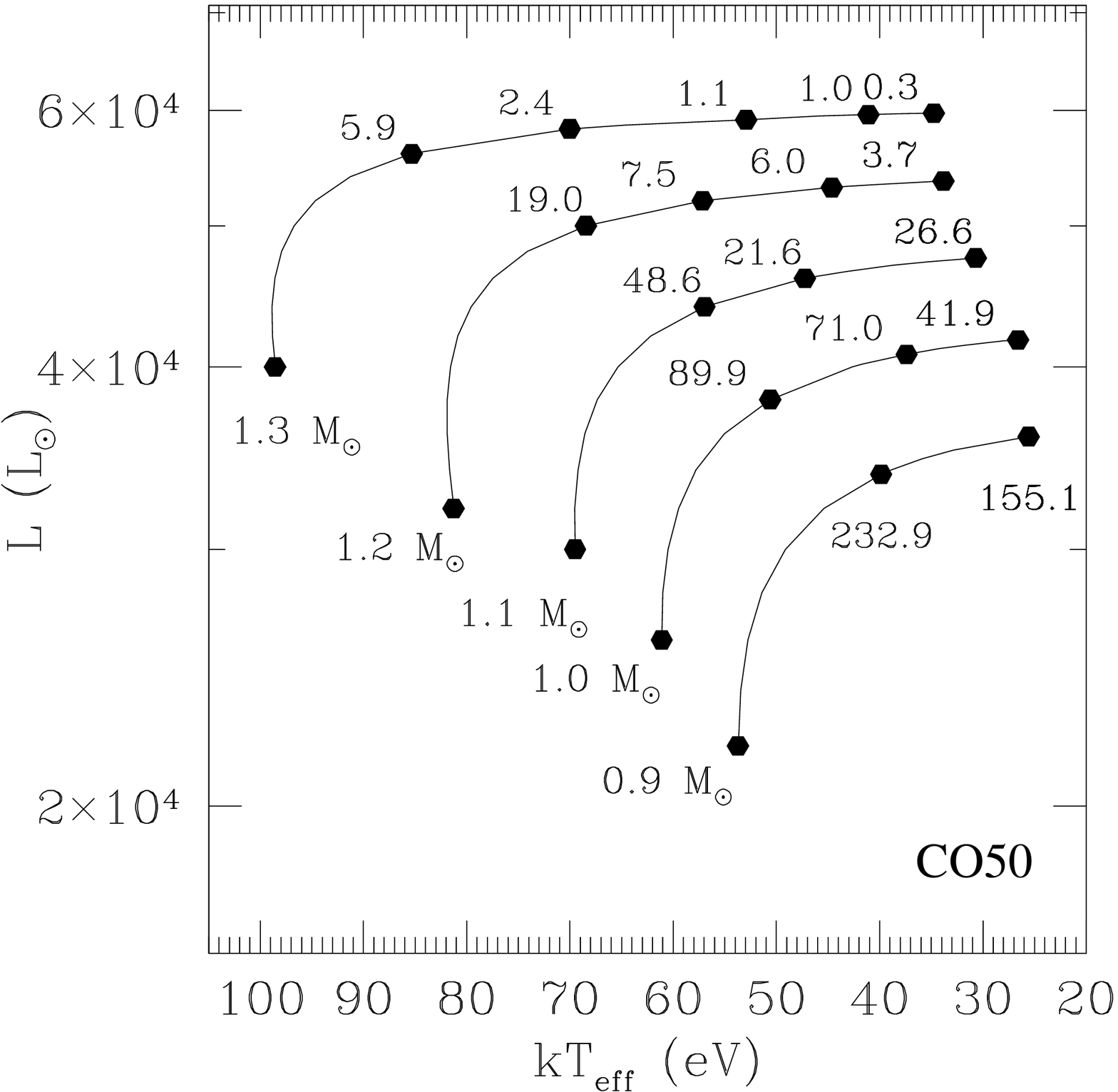}} 
\resizebox*{0.99\columnwidth}{!}{\includegraphics{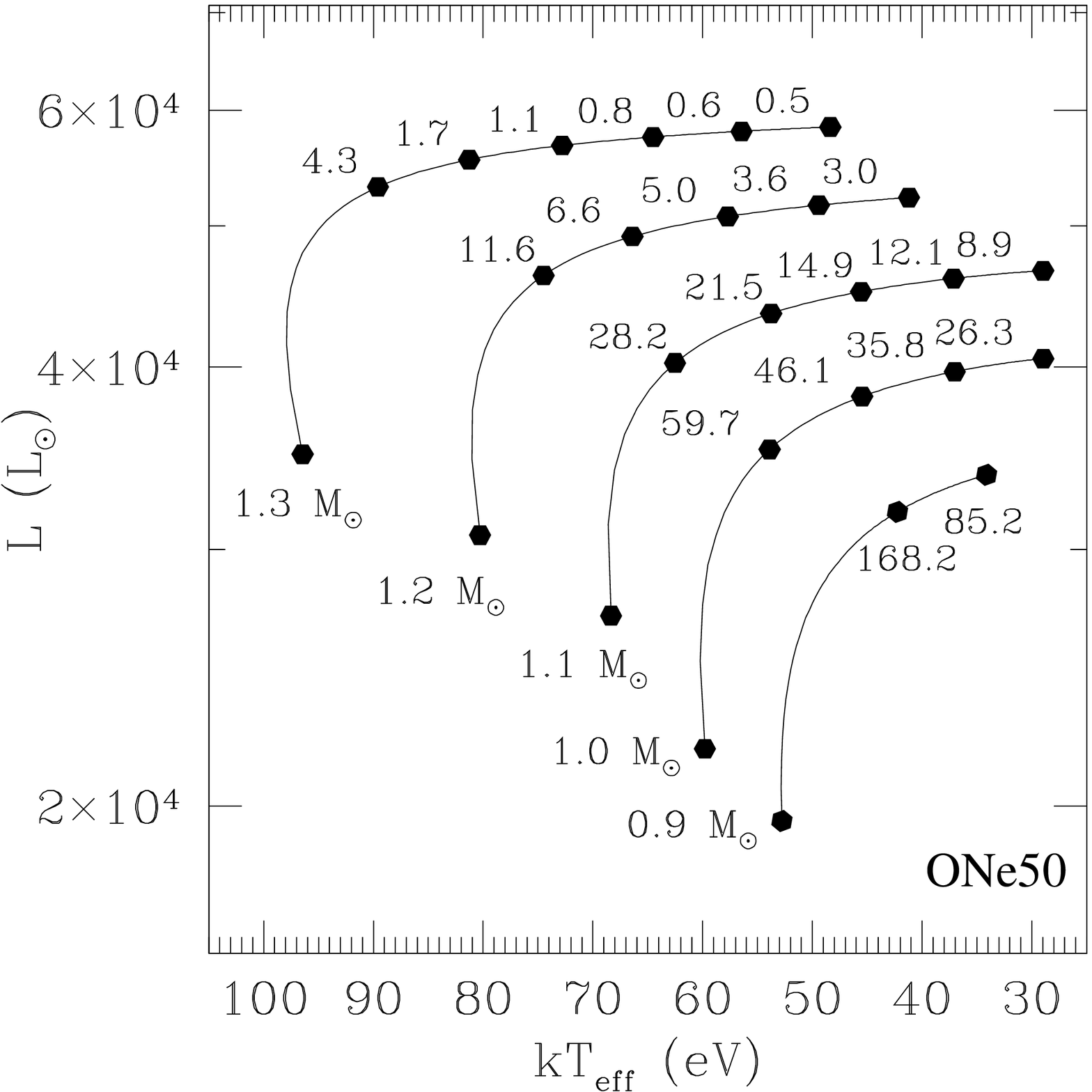}}
\caption{\label{tCO}Luminosity versus effective temperature for the 
high luminosity branch of CO50 and ONe50 envelopes. 
Time in days needed for the envelope to evolve 
between two adjacent ticks is indicated.}
\end{figure*}

\begin{figure*}
\resizebox*{0.99\columnwidth}{!}{\includegraphics{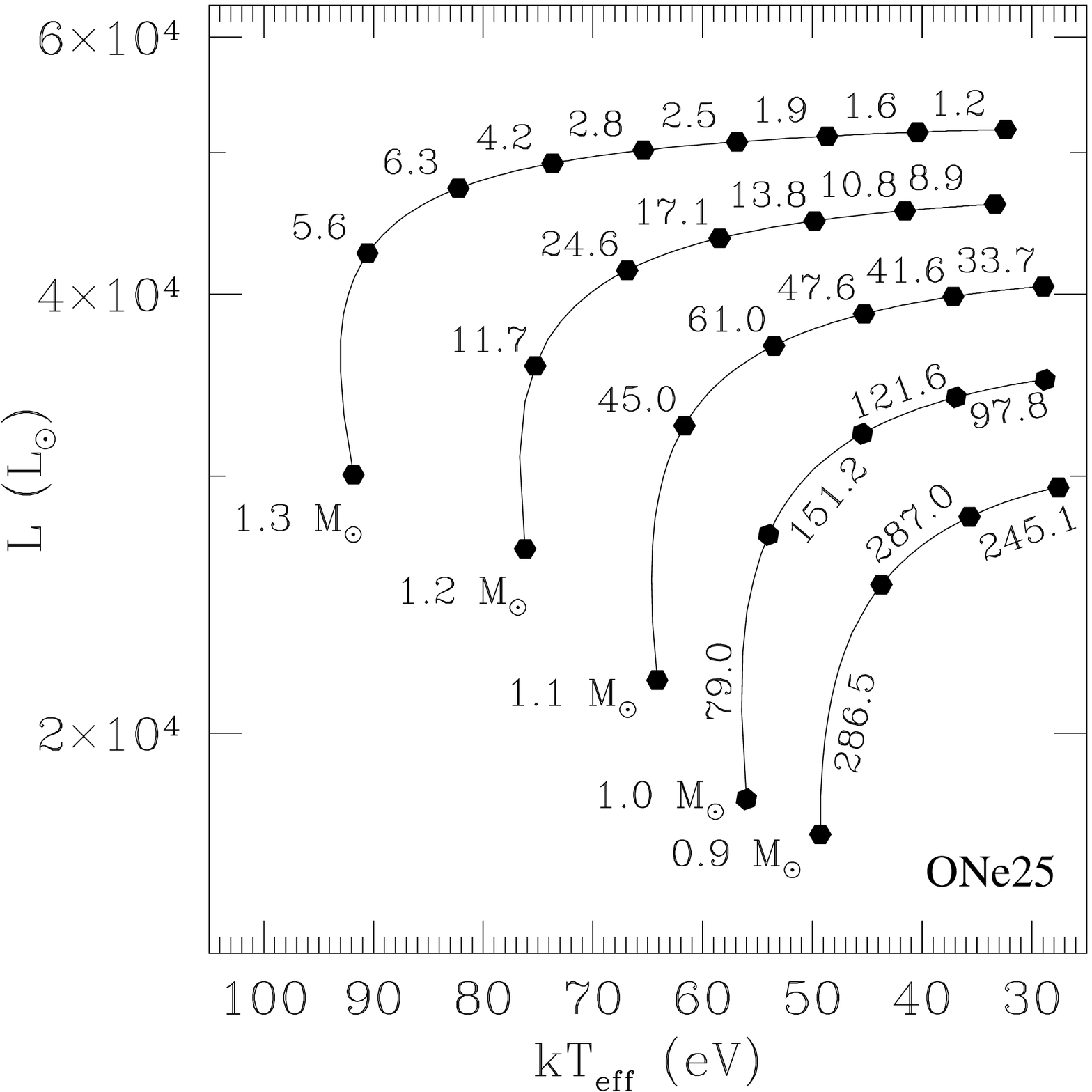}} 
\resizebox*{0.99\columnwidth}{!}{\includegraphics{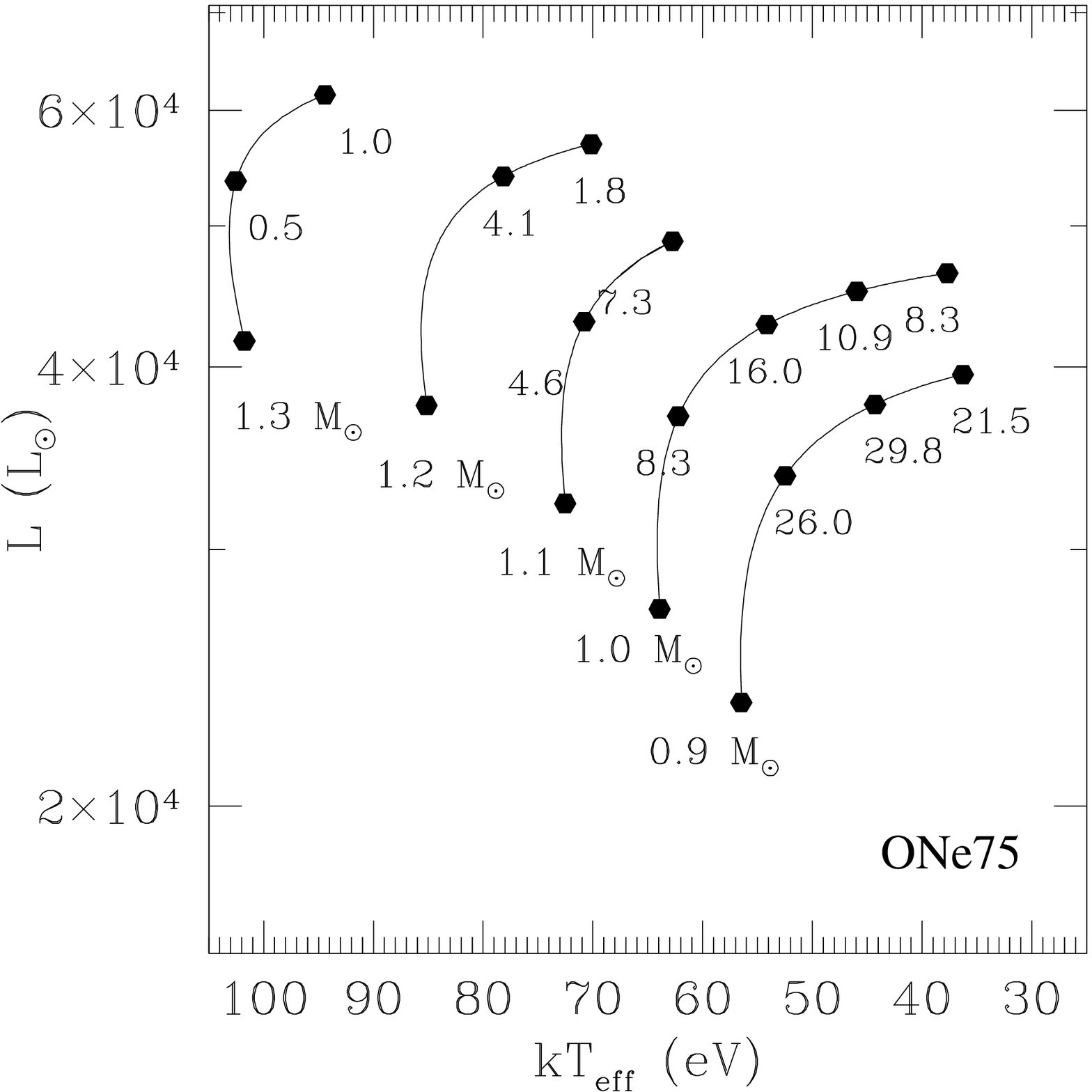}}
\caption{\label{tone2575}
Same as figure \ref{tCO} for ONe25 and ONe75 envelope models.}
\end{figure*}

An estimation of the evolution of a steady hydrogen burning white
dwarf envelope in the plateau of quasi-constant luminosity can be
approximated by a series of stationary envelope models with decreasing
envelope masses. The layers of the envelopes whose hydrogen has been
processed are assimilated to the core and do not take part in the
integration of the next envelope model (the increment in the core
mass is negligible). The time $\Delta$t needed for the envelope to
change its mass $\Delta$M$_{env}$ only due to hydrogen burning
can be estimated as 

\begin{equation}
\label{EQU:time_evaluation}
\Delta t=\epsilon \frac{\Delta M_{\rm env}X_{\rm H}}{L}
\end{equation}
\noindent
where X$_{\rm H}$ is the hydrogen mass fraction, L
is the mean luminosity in this period (which varies little in the
constant bolometric luminosity phase) and $\epsilon$ is the
energy released per gram of hydrogen processed ($\epsilon=5.98\times 10^{18}$erg g$^{-1}$).
In figures \ref{tCO}-\ref{tone2575},
time intervals needed for the envelope to evolve between two adjacent
marked points are indicated in days. Evolution proceeds slower for smaller white
dwarfs, with lower luminosities. Time intervals are also longer for
envelopes with increasing hydrogen abundance, which have smaller luminosities.

The times found in this work are much shorter than the evolutionary times
usually found in the literature for stationary hydrogen burning on
white dwarfs, including for instance the times found by Iben (\cite{ibe82}).
The difference can be explained by the smaller hydrogen fraction in
the envelope: the equilibrium configuration is hotter and brighter
for hydrogen poorer envelopes, and envelopes masses are smaller. To
keep this higher luminosity, hydrogen is consumed faster and, in addition,
for the same amount of hydrogen consumed, a larger fraction of the
envelope is reduced. As envelope masses are smaller and are reduced
at a faster rate, the total timescale to consume the envelope is shorter. 

These timescales can be compared with the nuclear-burning lifetime
for classical novae hydrostatic remnants found in the literature. 
In Starrfield (\cite{sta89}), this lifetime was estimated as
\begin{equation}
\tau_{\rm nuc}=400\left( \frac{M_{\rm H}}{10^{-4}M_{\odot }}\right) \left( \frac{L}{2\times 10^{4}L_{\odot }}\right) ^{-1} \rm yr
\end{equation}

where M$_{\rm H}$ is the hydrogen mass and L the total luminosity.
A similar expression, only a factor $\frac{5}{4}$ larger, was
given by MacDonald et al. (\cite{mac85}). In both cases, the nuclear lifetime
for an envelope of $10^{-5}-10^{-4}\rm M_{\odot}$ is of decades
or even centuries. According to hydrodynamical simulations, $10^{-5}-10^{-4}\rm M_{\odot}$
is the mass required for the thermonuclear runaway to be triggered
(Jos\'e \& Hernanz \cite{jh98}, Starrfield \cite{sta98}, Prialnik \& Kovetz \cite{pri95}), 
and the fraction of the accreted
mass that is expected to be ejected is not large enough as to reduce
the mass of the remnant by two orders of magnitude. But in the present
work, stable envelopes have been found to have masses of 
$\sim 10^{-6}\rm M_{\odot}$, which requires an extra mass
loss mechanism after the nova outburst for the envelope to reach the
equilibrium configuration. Once the envelope has reached the steady
hydrogen burning phase, the nuclear lifetime 
can be rewritten with our typical values (M$_{\rm env}=10^{-6}\rm M_{\odot}$,
L=4$\times 10^{4}\rm L_{\odot}$) as
\begin{equation}
\label{EQUnuc}
\tau _{\rm nuc}=2X_{\rm H}\left( \frac{M_{\rm env}}{10^{-6}M_{\odot }}\right) \left( \frac{L}{4\times 10^{4}L_{\odot }}\right) ^{-1} \rm yr
\end{equation}
For X=0.35 (as in ONe50 and CO50 models), $\tau _{\rm nuc}$=0.7 yr$\simeq$ 255 days,
which agrees with the times in figure \ref{tCO}.

An indication of the typical evolutionary times for our models is given in 
the last column in table \ref{tabsum}, which lists the time needed 
for the envelope to change its effective temperature
the last $\sim$ 10 eV before reaching the maximum effective
temperature.


\begin{table*}
\centering
\caption{\label{tabsum} Main properties of the envelope models.}

\begin{tabular}{ c c c c c c c c }
\hline \hline
\noalign{\smallskip}
Model & M$_{\rm c}$ & L$^{\rm plateau}$ & kT$^{\rm max}_{\rm eff}$ & T$_{\rm eff}^{\rm max}$ &
M$^{\rm max}_{\rm env}$ & M$^{\rm min}_{\rm env}$ & $\Delta$ t$^{(\rm a)}_{10, \rm eV}$ \\
 & (M$_{\odot})$ & ($10^{4}$L$_{\odot}$) & (eV) & ($10^{5}$K ) & ($10^{-6}$M$_{\odot}$) &
($10^{-6}$M$_{\odot}$) & (days)\\
\noalign{\smallskip}
\hline
\noalign{\smallskip}
\multicolumn{1}{c}{}&0.9&3.9&57&6.6&2.3&1.9&47\\
\multicolumn{1}{c}{ONe75}&1.0&4.4&64&7.4&1.4&1.2&24\\
\multicolumn{1}{c}{(X=0.18)}&1.1&5.2&73&8.5&0.84&0.71&12\\
\multicolumn{1}{c}{}&1.2&5.7&86&9.9&0.40&0.35&4.6\\
\multicolumn{1}{c}{}&1.3&6.2&103&11.9&0.16&0.15&1.1\\
\noalign{\smallskip}
\hline
\noalign{\smallskip}
\multicolumn{1}{c}{}&0.9&3.5&53&6.1&2.7&1.5&160\\
\multicolumn{1}{c}{ONe50}&1.0&4.1&60&7.0&1.7&1.2&78\\
\multicolumn{1}{c}{(X=0.35)}&1.1&4.7&69&8.0& 0.99&0.72&37\\
\multicolumn{1}{c}{}&1.2&5.2&81&9.4&0.45&0.35&14\\
\multicolumn{1}{c}{}&1.3&5.6&98&11.4&0.19&0.15&4.9\\
\noalign{\smallskip}
\hline
\noalign{\smallskip}
\multicolumn{1}{c}{}&0.9&2.9&49&5.7&3.0&2.2&430\\
\multicolumn{1}{c}{ONe25}&1.0&3.5&56&6.5&2.1&1.4&210\\
\multicolumn{1}{c}{(X=0.53)}&1.1&4.0&64&7.5&1.20&0.81&98\\
\multicolumn{1}{c}{}&1.2&4.6&77&8.9&0.60&0.39&36\\
\multicolumn{1}{c}{}&1.3&5.2&93&10.8&0.22&0.16&12\\
\noalign{\smallskip}
\hline
\noalign{\smallskip}
\multicolumn{1}{c}{}&0.9&3.6&54&6.2&2.9&1.9&230\\
\multicolumn{1}{c}{CO50}&1.0&4.2&61&7.1&1.8&1.2&90\\
\multicolumn{1}{c}{(X=0.35)}&1.1&4.7&70&8.1&1.0&0.72&48\\
\multicolumn{1}{c}{}&1.2&5.4&82&9.5&0.49&0.35&19\\
\multicolumn{1}{c}{}&1.3&6.0&99&14.7&0.19&0.15&6\\
\noalign{\smallskip}
\hline
\noalign{\smallskip}
\end{tabular}
\begin{list}{}{}
\item[$^{\rm {a}}$] Time needed for the envelope to evolve from kT$_{\rm eff}\simeq$kT$^{\rm max}_{\rm eff}-10$eV
to kT$^{\rm max}_{\rm eff}$.
\end{list}
\end{table*}

\section{Summary}

A numerical model has been developed to simulate the physical conditions
in the steady hydrogen burning envelope of a white dwarf after a nova
outburst. A grid of envelope models has been computed for five different
white dwarf masses, from 0.9 to 1.3 M$_{\odot}$, and four chemical
compositions, corresponding to realistic CO and ONe white dwarf envelopes,
with several degrees of mixing between the core material and the solar-like
accreted matter. The results show that there exists a maximum luminosity
and a maximum effective temperature for every white dwarf mass and
composition. The maximum effective temperature divides the series
of models into two branches: a stable branch of high, quasi-constant
luminosity, and a low-luminosity unstable branch, with constant photospheric
radius. For every white dwarf mass and composition there is also a
minimum envelope mass, which occurs in the vicinity of the maximum
effective temperature. The mass of the stable branch envelopes is
almost constant, decreasing only slowly for increasing effective temperatures.
The plateau luminosity, maximum effective temperature, and  maximum 
and minimum envelope masses for each model are summarized in table \ref{tabsum}, 
as well as the typical evolutionary timescales.
Evolutionary timescales are longer for less massive white dwarfs,
since luminosity is lower in these cases. The evolution also proceeds
slower for envelopes with increasing hydrogen abundances, for the same reason.

Each of the overall properties of the envelopes is related to the
total white dwarf mass and its chemical composition, and we have found analytical 
approximations for our ONe envelope models (equations \ref{equXLR}-\ref{equMmin}).
Our results show that masses of stable envelopes with steady hydrogen
burning ($\sim 10^{-6}\rm M_{\odot}$) are in general smaller by
at least one order of magnitude than envelope masses needed to trigger
the outburst ($\sim 10^{-5}\rm M_{\odot}$), according to hydrodynamical
models. Since those models predict that only a fraction of the accreted
envelope is ejected, it is clear that some mass loss mechanism must
act after the outburst, before the steady hydrogen burning can be
established in the envelope. An interesting possibility for the mass
loss is, as pointed out by Tuchman \& Truran (\cite{tuc98}), the dynamical
instabilities of the envelope itself when it is left after the outburst
with a mass much larger than that of stable configurations. 

With the masses found for stable envelopes, the nuclear time-scales
are of some weeks or months, similar to observed for post-outburst classical novae. 
For accreted envelopes highly mixed with the degenerate core, the short duration of 
the steady hydrogen burning phase could explain the lack of soft X-ray emission in most 
novae observed, even if not all the accreted envelope is ejected by the outburst. 
For those cases where the soft X-ray emission has been detected,
the photospheric properties obtained with the envelope models and
their temporal evolution can be directly compared with the observational results.
The comparison of observations with the present models has
been performed in Sala \& Hernanz (2005) 
for the ROSAT observations of V1974 Cyg. 
From this comparison, the white dwarf mass, and envelope mass and composition 
have been constrained.

The present envelope models represent a powerful tool for the study of 
classical novae, specially if their post-outburst phases are 
monitored in X-rays. With short and frequent observations, with a good 
sensitivity in the soft X-ray range and a moderate spectral resolution, 
the comparison with our envelope models makes possible the determination of the 
main parameters of the white dwarf. Our envelope models can also be applied to 
the study of other systems with stationary hydrogen burning on top of a white dwarf, 
(such as supersoft X-ray sources or symbiotic stars).

\begin{acknowledgements}
This research has been partially funded by the MEC project AYA2004-06290-C02-01. 
GS acknowledges a FPI grant from the MEC.
\end{acknowledgements}

\end{document}